\documentclass[onecolumn]{mn2e}  
\usepackage[dvips]{graphics}
\usepackage{epsfig}
\usepackage{epsf}
\usepackage[figuresleft]{rotating}
\begin{document}
%
% MYMACROS.TEX
% uses macros defined in doublespace.sty e.g \singlespace
% Stolen again from MJH Nov 26 1992, subsequent messes are all mine...
\newcommand{\getinptfile}
{\typein[\inptfile]{Input file name}
\input{\inptfile}}
% ================================================
% math mode macros
% force math mode

\newcommand{\mysummary}[2]{\noi {\bf SUMMARY}#1 \\ \noi \sl #2 \\ \capline 
	\hspace{-.13in} \raisebox{.0in}{$\sqcap$} \rm }  
\newcommand{\mycaption}[2]{\caption[#1]{\footnotesize #2}} % First argument:lof
\newcommand{\capline}{\mbox{}\hrulefill}
 % These \my... commands below create \junk.. everytime they're invoked;
 % \junk.. itself is called in mythesis.sty, in the \myheadings definitions;
 % one passes as an argument to \my... the running title of a section, say,
 % which appears in the headings. 
\newcommand{\mysection}[2]{ 
\section{\uppercase{\normalsize{\bf #1}}} \def\junksec{{#2}} } %
\newcommand{\mychapter}[2]{ \chapter{#1} \def\junkchap{{#2}}  % for big headers
\def\thesection{\arabic{chapter}.\arabic{section}}
\def\thesubsection{\thesection.\arabic{subsection}}
\def\thesubsubsection{\thesubsection.\arabic{subsubsection}}
\def\theequation{\arabic{chapter}.\arabic{equation}}
\def\thefigure{\arabic{chapter}.\arabic{figure}}
\def\thetable{\arabic{chapter}.\arabic{table}}
}
\newcommand{\mysubsection}[2]{ \subsection{#1} \def\junksubsec{{#2}} }
	% You can't mark these sections with \footnote; instead 
	% one types \thenote in the (first) bracket, and use \footnotetext
	% outside of the curly brackets to send text to the bottom of the page.
	% the counters are taken care of automatically. 
\def\thenote{\addtocounter{footnote}{1}$^{\scriptstyle{\arabic{footnote}}}$ }

\newcommand{\myfm}[1]{\mbox{$#1$}}
% produce <~ and >~
\def\spose#1{\hbox to 0pt{#1\hss}}	% Definition of .le., .ge. symbols
\def\ltabout{\mathrel{\spose{\lower 3pt\hbox{$\mathchar"218$}} % courtesy AJC1.
     \raise 2.0pt\hbox{$\mathchar"13C$}}}
\def\gtabout{\mathrel{\spose{\lower 3pt\hbox{$\mathchar"218$}}
     \raise 2.0pt\hbox{$\mathchar"13E$}}}
\newcommand{\ltsim}{\raisebox{-0.5ex}{$\;\stackrel{<}{\scriptstyle \backslash}\;$}}
\newcommand{\simlt}{\ltsim}
\newcommand{\simgt}{\gtsim}
%
% units in math mode in roman font
\newcommand{\unit}[1]{\ifmmode \:\mbox{\rm #1}\else \mbox{#1}\fi}
\newcommand{\ze}{\ifmmode \mbox{z=0}\else \mbox{$z=0$ }\fi }

%
% bold vector
\newcommand{\boldv}[1]{\ifmmode \mbox{\boldmath $ #1$} \else 
 \mbox{\boldmath $#1$} \fi}
%
% subscripts with roman font
\renewcommand{\sb}[1]{_{\rm #1}}%
% expectation value
\newcommand{\expec}[1]{\myfm{\left\langle #1 \right\rangle}}
\newcommand{\mone}{\myfm{^{-1}}}
\newcommand{\half}{\myfm{\frac{1}{2}}}
\newcommand{\nth}[1]{\myfm{#1^{\small th}}}
\newcommand{\ten}[1]{\myfm{\times 10^{#1}}}
	% Mathematical definitions / parameters used in the text
\newcommand{\abs}[1]{\mid\!\! #1 \!\!\mid}
\newcommand{\as}{a_{\ast}}
\newcommand{\asr}{(a_{\ast}^{2}-R_{\ast}^{2})}
\newcommand{\bvm}{\bv{m}}
\newcommand{\calf}{{\cal F}}
\newcommand{\calI}{{\cal I}}
\newcommand{\calm}{{v/c}}
\newcommand{\calminf}{{(v/c)_{\infty}}}
\newcommand{\calQ}{{\cal Q}}
\newcommand{\calR}{{\cal R}}
\newcommand{\calw}{{\it W}}
\newcommand{\co}{c_{o}}
\newcommand{\cs}{C_{\sigma}}
\newcommand{\cst}{\tilde{C}_{\sigma}}
\newcommand{\cv}{C_{v}}
\def\dbar{{\mathchar '26\mkern-9mud}}	
\newcommand{\deldelr}{\frac{\partial}{\partial r}}
\newcommand{\deldelR}{\frac{\partial}{\partial R}}
\newcommand{\deldeltheta}{\frac{\partial}{\partial \theta} }
\newcommand{\deldelphi}{\frac{\partial}{\partial \phi} }
\newcommand{\ddotrc}{\ddot{R}_{c}}
\newcommand{\ddotxc}{\ddot{x}_{c}}
\newcommand{\dotrc}{\dot{R}_{c}}
\newcommand{\dotxc}{\dot{x}_{c}}
\newcommand{\Estar}{E_{\ast}}
\newcommand{\grpsi}{\Psi_{\ast}^{\prime}}
\newcommand{\kboltz}{k_{\beta}}
\newcommand{\levi}[1]{\epsilon_{#1}}
\newcommand{\limaso}[1]{$#1 ( a_{\ast}\rightarrow 0)\ $}
\newcommand{\limasinfty}[1]{$#1 ( a_{\ast}\rightarrow \infty)\ $}
\newcommand{\limrinfty}[1]{$#1 ( R\rightarrow \infty,t)\ $}
\newcommand{\limro}[1]{$#1 ( R\rightarrow 0,t)\ $}
\newcommand{\limrso}[1]{$#1 (R_{\ast}\rightarrow 0)\ $}
\newcommand{\limxo}[1]{$#1 ( x\rightarrow 0,t)\ $}
\newcommand{\limxso}[1]{$#1 (\xs\rightarrow 0)\ $}
\newcommand{\ls}{l_{\ast}}
\newcommand{\Ls}{L_{\ast}}
\newcommand{\mean}[1]{<#1>}
\newcommand{\ms}{m_{\ast}}
\newcommand{\Ms}{M_{\ast}}
\def\nb{{\sl N}-body }
\def\nbt{{\sf NBODY2} }
\def\nb1{{\sf NBODY1} }
\newcommand{\nuoned}{\nu\sb{1d}}
\newcommand{\ra}{\rightarrow}
\newcommand{\Ra}{\Rightarrow}
\newcommand{\rc}{r_{c} } % (t)}
\newcommand{\Rc}{R_{c} } % (t)}
\newcommand{\res}[1]{{\rm O}(#1)}
\newcommand{\rnsa}{(r^{2}-a^{2})}
\newcommand{\Rnsa}{(R^{2}-a^{2})}
\newcommand{\rs}{r_{\ast}}
\newcommand{\Rs}{R_{\ast}}
\newcommand{\Rsa}{(R_{\ast}^{2}-a_{\ast}^{2})}
\newcommand{\sa}{\sigma } % [R,t]}
\newcommand{\sac}{\sigma_{c} } % [t]}
\newcommand{\sas}{\sigma_{\ast} } % [R_{\ast}]}
\newcommand{\sasp}{\sigma^{\prime}_{\ast}}
\newcommand{\saxs}{\sigma_{\ast} } % [x_{\ast}]}
\newcommand{\sech}{{\rm sech}}
\newcommand{\tff}{t\sb{ff}} 
\newcommand{\ti}{\tilde}
\newcommand{\trel}{t\sb{rel}}
\newcommand{\ts}{\tilde{\sigma} } % [a,t]}
\newcommand{\tss}{\tilde{\sigma}_{\ast} } % [a_{\ast}]}
\newcommand{\vcol}{v\sb{col}}
\newcommand{\vs}{v_{\ast}  } % [R_{\ast}]}
\newcommand{\vsp}{v^{\prime}_{\ast}}
\newcommand{\vxs}{v_{\ast}  } % [x_{\ast}]}
\newcommand{\xs}{x_{\ast}}
\newcommand{\xc}{x_{c} } % [t]}
\newcommand{\xistar}{\xi_{\ast}}
\newcommand{\rmd}{\ifmmode \:\mbox{{\rm d}}\else \mbox{ d}\fi }
\newcommand{\rmD}{\ifmmode \:\mbox{{\rm D}}\else \mbox{ D}\fi }
\newcommand{\valfven}{v_{{\rm Alfv\acute{e}n}}}

%
% ================================================
%  Abreviations used  in the text ... 
\newcommand{\noi}{\noindent}
\newcommand{\bc}{boundary condition }
\newcommand{\bcs}{boundary conditions }
\newcommand{\Bcs}{Boundary conditions }
\newcommand{\lhs}{left-hand side }
\newcommand{\rhs}{right-hand side }
\newcommand{\wrt}{with respect to }
\newcommand{\iras}{{\sl IRAS }}
\newcommand{\cobe}{{\sl COBE }}
\newcommand{\Oh}{\myfm{\Omega h}}
% \newcommand{\vs}{\vspace{3mm}}
%
% Latin expressions, etc ... 
\newcommand{\etal}{{\em et al.\/ }}
\newcommand{\eg}{{\em e.g.\/ }}
\newcommand{\etc}{{\em etc.\/ }}
\newcommand{\ie}{{\em i.e.\/ }}
\newcommand{\viz}{{\em viz.\/ }}
\newcommand{\cf}{{\em cf.\/ }}
\newcommand{\via}{{\em via\/ }}
\newcommand{\apriori}{{\em a priori\/ }}
\newcommand{\adhoc}{{\em ad hoc\/ }}
\newcommand{\viceversa}{{\em vice versa\/ }}
\newcommand{\versus}{{\em versus\/ }}
\newcommand{\qed}{{\em q.e.d. \/}}
\newcommand{\<}{\thinspace}
%
%%%%%%%%%%%%%%%%%%%%%%%%%%%%%%%%%%%%%%%%%%%%%%%%%%%%%%%%%%%%%%%%%%%%%%
% Astro stuff
%%%%%%%%%%%%%%%%%%%%%%%%%%%%%%%%%%%%%%%%%%%%%%%%%%%%%%%%%%%%%%%%%%%%%%
% units
\newcommand{\km}{\unit{km}}
\newcommand{\kms}{\unit{km~s\mone}}
\newcommand{\kmsa}{\unit{km~s\mone~arcmin}}
\newcommand{\kpc}{\unit{kpc}}
\newcommand{\mpc}{\unit{Mpc}}
\newcommand{\hkpc}{\myfm{h\mone}\kpc}
\newcommand{\hmpc}{\myfm{h\mone}\mpc}
\newcommand{\parsec}{\unit{pc}}
\newcommand{\cm}{\unit{cm}}
\newcommand{\yr}{\unit{yr}}
\newcommand{\au}{\unit{A.U.}}
\newcommand{\AU}{\au}
\newcommand{\gm}{\unit{g}}
\newcommand{\solarm}{\unit{M\sun}}
\newcommand{\Lsun}{\unit{L\sun}}
\newcommand{\Rsun}{\unit{R\sun}}
\newcommand{\seconds}{\unit{s}}
\newcommand{\micro}{\myfm{\mu}}
\newcommand{\Mdot}{\myfm{\dot M}}
%
%Spherical coords
% these macros write the deg arcmin or arcsec symbols with periods aligned.
% e.g. 3\sec5 produces 3".5 with " above the period.
% (For some reason \hspace doesn't work with ' or ").
% (\stackrel leaves too much space and shifts the "." up off the baseline) 
% NB: \dgr gives degrees (\deg is a TeX command)
% 
%
\newcommand{\dgr}{\myfm{^\circ} }
\newcommand{\ddgr}{\mbox{\dgr\hskip-0.3em .}}
\newcommand{\mnt}{\mbox{\myfm{'}\hskip-0.3em .}}
\newcommand{\scnd}{\mbox{\myfm{''}\hskip-0.3em .}}
\newcommand{\hr}{\myfm{^{\rm h}}}
\newcommand{\dhr}{\mbox{\hr\hskip-0.3em .}}
% what about ^m
%
%\newcommand{\ra}{\myfm{\alpha}}
%\newcommand{\dec}{\myfm{\delta}}
%\newcommand{\radec}[2]{\myfm{\alpha = #1\hr}, \myfm{\delta = #2\dgr}}
%\newcommand{\dradec}[4]{\myfm{\alpha = #1\dhr#2}, \myfm{\delta = #2\ddgr#4}}
%
%\newcommand{\xyz}[3]{($#1$, $#2$, $#3$)}
%
% magnitudes (\mag is a TeX command).
%\newcommand{\mg}{\myfm{^{\rm m}}}
% decimal magnitudes
%\newcommand{\dmg}{\mbox{\mg\hskip-0.3em .}}
%
%%%%%%%%%%%%%%%%%%%%%%%%%%%%%%%%%%%%%%%%%%%%%%%%%%%%%%%%%%%%%%%%%%%%%%%
% References
%%%%%%%%%%%%%%%%%%%%%%%%%%%%%%%%%%%%%%%%%%%%%%%%%%%%%%%%%%%%%%%%%%%%%%%
%
	% New environment introduced to cross-reference figures.
%\newcounter{fig}
%\newcommand{\fig}[1]
%   {\thefig\newcounter{#1}
%    \setcounter{#1}{\thefig}\addtocounter{fig}{1}\hspace{-12pt} } % 12pt,10pt
%\newenvironment{figures}[1]{ {\bf Figure #1.}}{}
%
%
\newcommand{\refindent}{\par\noindent\hangindent=0.5in\hangafter=1}
\newcommand{\figpar}{\par\noindent\hangindent=0.7in\hangafter=1}
%
% biblography command  nicked from MNRAS.STY but heavily modified
% usage: \begin{thebibliography} 
%                          \item Lynden-Bell...,  
%                          \item Lahav etal ...
%        \end{thebibliography}
% Modified it to use single spacing (from doublespace.sty)
%

\newcommand{\mybiblio}{\vspace{1cm}
		       \setcounter{subsection}{0}
		       \addtocounter{section}{1}
		       \def\junksec{References} 
 }

%\renewenvironment{thebibliography}
%{  \section*{\\ \\ \uppercase{References}}   
% \small
% \begin{singlespace}
% \begin{list}{}{\setlength{\labelsep}{3pt}
%		\setlength{\leftmargin}{0.5cm}
%                \setlength{\rightmargin}{0.0cm}
%                \setlength{\itemindent}{-0.5cm}
%                \setlength{\itemsep}{0.25\baselineskip}
%                \setlength{\parsep}{0cm}  % space between lines
%                \def\newblock{\hskip .11em plus .33em minus -.07em}
%\sloppy
%                \sfcode`\.=1000\relax % does not treat ``.'' as a period
%               } % end of list
%} % thebibliography begargs
%{\end{list}
% \end{singlespace}
%
%  \addcontentsline{toc}{section}{References}
%
%}%thebibliography endargs
% end of Citations 
%

% Reference formats:
% MN  format has Arp, A., Burbidge, B. & C, C., 1991. Astrophys. J., 301, 1.
% ApJ format has Arp, A., Burbidge, B., & C, C. 1991, ApJ, 301, 1
% A&A format has Arp, A., Burbidge, B., C, C., 1991, ApJ 301, 1
% 
% we adopt       Arp, A., Burbidge, B. & C, C. 1991. Astrophys. J., 301, 1.
% ie like MN but without the , before the year although A&A format is logical
%
% ---------------------------------------------------------------------------
% MNRAS-style reference macros. 
%
% example:
% To generate reference to a paper in Ap.J. volume 301, p.1
% write \apj{Arp, A., Burbidge, B. & C, C., 1990}{300}{123}
%
% use \item for full reference
% 
	% References: list of journals.

\newcommand{\vol}[2]{ {\bf#1}, #2}
\newcommand{\jour}[4]{#1. {\it #2\/}, {\bf#3}, #4}
\newcommand{\physrevd}[3]{\jour{#1}{Phys Rev D}{#2}{#3}}
\newcommand{\physrevlett}[3]{\jour{#1}{Phys Rev Lett}{#2}{#3}}
\newcommand{\aaa}[3]{\jour{#1}{A\&A}{#2}{#3}}
\newcommand{\aaarev}[3]{\jour{#1}{A\&A Review}{#2}{#3}}
\newcommand{\aaas}[3]{\jour{#1}{A\&A Supp.}{#2}{#3}}
\newcommand{\aj}[3]{\jour{#1}{AJ}{#2}{#3}}
\newcommand{\apj}[3]{\jour{#1}{ApJ}{#2}{#3}}
\newcommand{\apjl}[3]{\jour{#1}{ApJ Lett.}{#2}{#3}}
\newcommand{\apjs}[3]{\jour{#1}{ApJ Suppl.}{#2}{#3}}
\newcommand{\araa}[3]{\jour{#1}{ARAA}{#2}{#3}}
\newcommand{\mn}[3]{\jour{#1}{MNRAS}{#2}{#3}}
\newcommand{\mnras}{\mn}
\newcommand{\jgeo}[3]{\jour{#1}{Journal of Geophysical Research}{#2}{#3}}
\newcommand{\qjras}[3]{\jour{#1}{QJRAS}{#2}{#3}}
\newcommand{\nat}[3]{\jour{#1}{Nature}{#2}{#3}}
\newcommand{\pasa}[3]{\jour{#1}{PAS Australia}{#2}{#3}}
\newcommand{\pasj}[3]{\jour{#1}{PAS Japan}{#2}{#3}}
\newcommand{\pasp}[3]{\jour{#1}{PAS Pacific}{#2}{#3}}
\newcommand{\rmp}[3]{\jour{#1}{Rev. Mod. Phys.}{#2}{#3}}
\newcommand{\science}[3]{\jour{#1}{Science}{#2}{#3}}
\newcommand{\vistas}[3]{\jour{#1}{Vistas in Astronomy}{#2}{#3}}

%
%
% Some definitions I use in these instructions.
%
\newcommand{\leftb}{<\!\!} \newcommand{\rightb}{\!\!>}
\newcommand{\oversim}[2]{\protect{\mbox{\lower0.5ex\vbox{%
  \baselineskip=0pt\lineskip=0.2ex
  \ialign{$\mathsurround=0pt #1\hfil##\hfil$\crcr#2\crcr\sim\crcr}}}}} 
\newcommand{\simgreat}{\mbox{$\,\mathrel{\mathpalette\oversim>}\,$}} % >~ sign
\newcommand{\simless} {\mbox{$\,\mathrel{\mathpalette\oversim<}\,$}} % <~ sign
%
%
% Page-setting commands 
%\marginparwidth 1.25in
%\marginparsep .125in
%\marginparpush .25in
%\reversemarginpar
%
%
%%%%%%%%%%%%%%%%%%%%%%%%%%%%%%%%%%%%%%%%%%%%%%%%%%%%%%%%%%%%%%%%%%%%%%%%
\title{The dynamical evolution of Taurus--Auriga-type aggregates}
   \author[Kroupa \& Bouvier]{Pavel Kroupa$^{1,2}$, Jerome Bouvier$^1$\\ 
$^1$Laboratoire d'Astrophysique de l'Observatoire de Grenoble, BP 53,
   F-38041 Grenoble Cedex 9, France \\
$^2$Institut f\"ur Theoretische Physik und Astrophysik der
   Universit\"at Kiel, D-24098 Kiel, Germany}

\maketitle

\begin{abstract} 
Star formation in the Taurus--Auriga (TA) molecular clouds is
producing binary-rich aggregates containing at most a few dozen
systems within a region spanning one~pc without massive stars. This
environment is very different to another well-studied star-forming
event which produced the Orion Nebula cluster (ONC). The ONC contains
a few thousand systems within a region of one~pc including massive
stars. Differences between these two environments have been
found. Notably, the ONC has a significantly smaller binary proportion
but a significantly larger number of isolated brown dwarfs (BDs) per
star than TA. The aim of the present project is to investigate if
these differences can be explained through stellar-dynamical evolution
alone. The stellar-dynamical issue is very relevant because dense
environments destroy binaries liberating BD companions, possibly
leading to the observed difference between the TA and ONC populations.
Here a series of high-precision $N-$body models of TA-like embedded
aggregates are presented, assuming the standard reference
star-formation model for the input populations according to which
stars and BDs form with the same kinematical, spatial and binary
properties.  After a discussion of the general evolution of the
aggregates, it is shown that the binary population indeed remains
mostly unevolved. Therefore, TA-type star formation cannot have added
significantly to the Galactic-field population. The standard model
leads to BDs tracing the stellar distribution, apart from a
high-velocity tail ($v\simgreat 1$~km/s) which leads to a more widely
distributed spatial distribution of single BDs.  The slow-moving BDs,
however, retain a high binary proportion, this being an important
observational diagnostic for testing against the embryo-ejection
hypothesis.  Inferences about the IMF and the binary-star orbital
distribution functions are made in two accompanying papers with useful
implications for star formation and the possible origin of BDs.

{\keywords stars: formation --
stars: low-mass, brown dwarfs -- binaries: general
-- open clusters and associations: general -- Galaxy: stellar content
-- stellar dynamics}

\end{abstract}

%%%%%%%%%%%%%%%%%%%%%%%%%%%%%%%%%%%%%%%%%%%%%%%%%%%%%%%%%%%%%%%%%%%%%%%%%%%
\section{Introduction}
\label{sec:intro}

Stars appear never to form in true isolation, but mostly with partners
which are typically clustered into groups. These contain from a few
dozen stars, such as the aggregates in the TA star formation complex
(Gomez et al. 1993; Briceno et al. 2002; Hartmann 2002), through to
thousands of stars in rich clusters, such as the ONC (Hillenbrand \&
Hartmann 1998; Muench et al. 2002) to hundreds of thousands of stars
in very rich and massive young globular cluster-type assemblages. By
comparing the properties of stellar populations in such different
environments we learn how different physical conditions determine the
star-formation products.  Owing to their proximity and similar age
(about 1~Myr) the stellar groups in TA and the ONC are prime
candidates for detailed inter-comparisons.

Observational campaigns have established that most stars in TA are in
binary systems, while in the ONC the binary properties of low-mass
stars appear to be similar to those of the Galactic-field population
(Duch\^ene 1999).  Stars more massive than about $2\,M_\odot$ have not
formed in TA, while the ONC contains a number of O~stars. The ONC has
also been shown to contain many brown dwarfs (BDs) (Hillenbrand \&
Carpenter 2000; Muench et al. 2002), while there is a deficit of
sub-stellar objects in TA (Briceno et al. 2002).  Furthermore, the
stellar rotation distributions in the two star-forming regions are
intrinsically different (Clarke \& Bouvier 2000).  The two physically
different systems thus appear to produce different stellar
populations.  However, closer inspection may indicate otherwise. Thus,
massive stars cannot form in TA because the cloud cores are not
massive enough. This does not necessarily imply that the stellar
initial mass function (IMF) has a different form. It merely means that
the most massive stars can only form if sufficient material is
available.  The higher frequency of binary stars in TA may be due to
the TA aggregates being dynamically unevolved, given that the crossing
time through an aggregate is about equal to the age of the stars or
longer, while a typical star has crossed the ONC several times within
its life-time. This would have reduced a TA binary frequency to the
observed ONC level (Kroupa, Aarseth \& Hurley 2001, hereinafter KAH),
thus eroding the argument for different initial period
distributions. Similarly, close encounters may have removed
disk-breaking in the ONC leading to faster rotators, or there is a
systematic difference in the observed stellar mass ranges in the two
samples together with a mass-rotation dependence (Clarke \& Bouvier
2000).

The recent affirmation by Briceno et al. (2002) that TA contains
significantly fewer BDs per star than the ONC is an important clue
that may indeed constitute the first direct evidence for a variation
of the IMF, at least in the sub-stellar regime. The problem we are
faced with in view of this empirical evidence, which implies a
variation of a factor of two or so in the number of BDs per star, is
that the surplus BDs detected in the ONC may be low-mass companions
that have left their primaries due to encounters. This possibility, if
confirmed, would again spoil the claim for a different sub-stellar IMF
in TA.

To address this particular problem a number of stellar-dynamical
models of embedded TA-like aggregates are calculated using the same
methodology and assumptions as in a previous recent detailed study of
the dynamical evolution of the ONC and the Pleiades cluster. This
study by KAH finds rather excellent agreement with all available
observational constraints (radial density profile, kinematics, IMF and
binary properties for the ONC and the Pleiades) under a set of
assumptions that define the outcome of star formation. These
assumptions, which we refer to as the {\it standard reference
star-formation model} (or just ``standard model''), have previously
been found to lead to an excellent agreement with the Galactic field
population if the majority of stars form in small embedded clusters
containing not more than a few hundred binaries (Kroupa 1995,
hereinafter K2).

The calculations are performed with the Aarseth-{\sc Nbody6} variant
{\sc GasEx} described by KAH in their study of cluster formation. {\sc
GasEx} is a direct, high-precision $N-$body integrator which allows
accurate treatment of close encounters and multiple stellar systems in
clusters through special mathematical transformation techniques of the
equations of motion (regularisation).  Force-softening is not
implemented.  It is identical to Aarseth's {\sc Nbody6} code apart
from having additional input and output features and an additional
routine for setting-up a realistic initial binary-star
population. {\sc GasEx} also allows the inclusion of a time-varying
analytic background potential to model gas expulsion form an embedded
cluster, and incorporates a local Galactic-tidal field into the
equations of motion of each star. A comparison of the effect of
gas-expulsion on an embedded cluster using hydrodynamical computations
and $N-$body calculations assuming time-varying analytic background
potentials leads to the same results (Geyer \& Burkert 2001).

The purpose of this paper is to introduce the models and describe the
evolution of TA--like aggregates thereby focusing attention on their
BD members.  Companion papers (Kroupa et al. 2003; Kroupa \& Bouvier
2003) address, respectively, the (related) issues of IMF variation and
binary systems in TA, and the implications this has for the origin of
BDs and free-floating planetary mass objects.  This paper begins with
a description of the standard model (\S~\ref{sec:standmod}). The
TA--like models studied here are introduced in \S~\ref{sec:mods}, and
their evolution is described in \S~\ref{sec:ev}. The evolution of the
binary systems is studied in \S~\ref{sec:bins}, and \S~\ref{sec:distr}
discusses how the relative distribution of the BDs and stars can be
used to differentiate between formation models of BDs.  The
conclusions follow in \S~\ref{sec:concs}.

%%%%%%%%%%%%%%%%%%%%%%%%%%%%%%%%%%%%%%%%%%%%%%%%%%%%%%%%%%%%%%%%%%%%%%%%%%
\section{The standard reference star-formation model}
\label{sec:standmod}

A brief description of the standard model is presented here. It is
useful as standardised and realistic initial conditions for $N-$body
computations of star clusters.  The standard model is defined by a
minimal set of assumptions based on empirical and theoretical evidence
that describe the outcome of star formation. The model has been
developed in K2 to find the dominant star-formation events that
produced the Galactic field population, taking as an initial boundary
condition the observed pre-main sequence binary-star properties in
TA. It accounts for the properties of short-period binary systems, but
does not incorporate BDs. In the strict form, it therefore only
applies to late-type stars with masses in the approximate range
$0.08-1\,M_\odot$. This model leads to stellar populations in good
agreement with available observational data for Galactic-field
main-sequence stars and pre-main sequence stars in dense clusters
including the Pleiades (K2; Kroupa, Petr \& McCaughrean 1999; KAH).

The standard model can be used to search for variations of the IMF or
binary-star properties in dependence of star-formation conditions.  If
a population is found which has an abnormal IMF or unusual binary-star
properties, and if dynamical and stellar evolution cannot reproduce
these observations given the standard model, then a very strong case
for a variation of the IMF or binary-star properties will have been
found.  An example of such an application is provided by Kroupa
(2001).  Thus, if a modelled population is found to deviate
significantly from an observed population, then this would imply that
the initial conditions for that population deviate from the standard
model and thus that there is a variation of the star-formation
products since the standard model can reproduce a number of
populations with very different ages.

We note that while the available results and notably the visualisation
of pure hydrodynamical collapse computations are spectacular (Bate
2003), it is not attempted to use these results as input for the
$N-$body calculations. Feedback from the forming stars via outflows
and radiation will probably significantly change the outcome of the
numerical hydrodynamical experiments. These allow most of the gas to
accrete, while in reality star-formation reaches an efficiency of less
than 40~per cent (Lada 1999; Matzner \& McKee 2000).  In particular,
it is noted that the modern collapse computations that lack feedback
and radiative energy transport as well as magnetic forces lead to
important lessons about the gravitational processes acting during the
onset of star formation, but they are likely to form groups of
``stars'' that are too compact and therewith too dynamically
violent. Stellar feedback is also likely to trigger the formation of
other stars nearby, which again is not captured by the existing cloud
collapse computations. It is not clear how feedback can be included in
a self-consistent manner in such collapse computations, unless some
parametrisation for example of the collimated outflows is attempted,
and because full radiation transfer poses prohibitive challenges on
the computational requirements. It is for this reason that the
$N-$body approach to young embedded clusters will remain a major
work-horse in the future.

The philosophy followed here is to approach the outcome of star
formation from the ``other side'' by studying how well-defined
physical processes (mostly stellar-dynamics) affect binary-star
properties and stellar distributions.  Our ansatz with an analytical
time-varying potential (\S~\ref{sec:mods}) assumes the newly formed
stars to have decoupled hydrodynamically from their gaseous
environment and captures the most important physical aspect of the
problem, namely the initial confinement of the stellar population in
their embedded cluster (Geyer \& Burkert 2001).  By this approach we
infer the properties the stellar distribution functions should have at
the point in time when the dynamics becomes dominated by point-mass
gravitational interactions, and which should ultimately come out of
the cloud-collapse plus feedback work that will hopefully become
available in the (distant) future.

The standard model assumes 

\begin{enumerate}

\item \label{p1} All stars are paired randomly from the IMF to form
binary systems with primary mass $m_p$ and companion or secondary mass
$m_s \le m_p$.

\item \label{p2} The distribution of orbital elements (period,
eccentricity and mass ratio) does not depend on the mass of the
primary star, but allowance for eigenevolution (see below) is made. 

\item \label{p3} Stellar masses are not correlated with the
phase-space variables (no initial mass segregation in a cluster).

\end{enumerate}

Assumption~\ref{p1} leads to a flat initial mass-ratio distribution
for late-type primaries, $f_{\rm q}$, (fig.~12 in K2; fig.~5 in Malkov
\& Zinnecker 2001), and is consistent with the flat mass-ratio
distribution for $q\equiv m_2/m_1 \simgreat 0.2$ derived from
observational data of pre-main sequence binaries by Woitas, Leinert \&
K\"ohler (2001). They state that ``these findings are in line with the
assumption that for most multiple systems in T~associations the
components' masses are principally determined by fragmentation during
formation and not by the following accretion processes''. This in turn
is supported by the finding that the mass function of pre-stellar
cores in $\rho$~Oph already has the same shape as the Galactic-field
IMF thus indicating that the fragmentation of a molecular cloud core
defines the distribution of stellar masses (Motte, Andr\'e \& N\'eri
1998; Bontemps et al. 2001; Matzner \& McKee 2000). The above
assumptions include implicitly that any initial multiple proto-stellar
systems have decayed within a few crossing times (typically $10^4$~yr,
Reipurth 2000) to form the binary-rich initial population seen in the
about 1~Myr old populations studied here.

By extending the standard model to include BDs we change the
stellar pairing properties by allowing stars to have BD
companions. The fraction of such systems may be appreciable but
depends on the IMF for BDs. It is the purpose of the present study to
investigate the consequences of the extension.  Likewise, extension of
the standard model to massive stars implies that most O~stars will
have low-mass companions, but here we are not concerned with massive
stars.

Assumption~\ref{p2} is posed given the indistinguishable period
distribution function of Galactic-field G-dwarf, K-dwarf and M-dwarf
binary systems (fig.~7 in K2). The discordant period distributions
between the pre-main sequence binaries and the Galactic-field systems
can be nicely explained by disruption of wide-period binaries in small
embedded clusters containing a few hundred stars. This destruction
process also leads to the observed mass-ratio distribution for G-dwarf
primaries in the Galactic field. The model is also in good agreement
with the observed smaller binary fraction of M~dwarfs than of K~dwarfs
and G~dwarfs (K2).

Assumption~\ref{p3} allows investigation of the important issue
whether massive stars need to form at the centres of their embedded
clusters to explain the observed mass segregation in very young
clusters such as the ONC (Hillenbrand 1997; Muench et
al. 2003). Assumption~\ref{p3} is motivated by observations which
indicate that at least some massive stars appear to be surrounded by
massive disks suggesting growth of the massive star by disk accretion
rather than through coagulation of proto-stars (Figueredo et al. 2002;
Cesaroni et al. 2003), and by the observations that forming embedded
clusters are typically heavily sub-clustered, with massive stars
forming at various locations (e.g. Motte, Schilke \& Lis
2002). Cesaroni et al. (2003) find the massive proto-stars in the
forming cluster G24.78+0.08 to lie at various locations within the
central region with a radius of 0.5~pc, a region that also appears to
be heavily sub-structured. On-going $N-$body work is addressing the
issue if dynamical mass-segregation can account for the observed mass
segregation in the ONC for example, in which sub-structure has been
erased already (Scally \& Clarke 2002).  Available state-of-the art
work by Bonnell \& Davies (1998) suggests that in the absence of
initial subclustering two-body relaxation may not be able to account
for the observed mass segregation. This work is based on Aarseth's
{\sc Nbody2} code which requires force softening to treat close
stellar encounters, and more recent attempts with {\sc GasEx}
indicates that two-body relaxation may lead to the observed mass
segregation if the initial configuration of the binary-rich cluster is
very compact (Kroupa 2002). This topic needs to be studied in more
detail.  The alternative scenario is that coagulation of forming
proto-stars in the densest embedded cluster region with continued
accretion of low-angular momentum material onto the forming cluster
core leads to the build-up of a core of massive stars there (Bonnell,
Bate \& Zinnecker 1998; Klessen 2001), which is tentatively supported
by observational data for very young clusters and older open clusters
(Raboud \& Mermilliod 1998).

The initial distribution functions that are needed to describe a
stellar population are the IMF, the period and eccentricity
distribution functions. The IMF is conveniently (for computational
purposes) taken to be a multi-power-law form,
\begin{equation}
\xi (m) = k\left\{
          \begin{array}{l@{\quad\quad,\quad}l}
   \left({m\over m_{\rm H}}\right)^{-\alpha_0}  &m_l < m \le m_{\rm H}, \\
   \left({m\over m_{\rm H}}\right)^{-\alpha_1}  &m_{\rm H} < m \le m_0, \\
   \left[\left({m_0\over m_{\rm H}}\right)^{-\alpha_1}\right] 
        \left({m\over m_0}\right)^{-\alpha_2} 
        &m_0 < m \le m_1,\\
   \left[\left({m_0\over m_{\rm H}}\right)^{-\alpha_1}
        \left({m_1\over m_0}\right)^{-\alpha_2}\right] 
        \left({m\over m_1}\right)^{-\alpha_3} 
        &m_1 < m \le m_2,\\
   \left[\left({m_0\over m_{\rm H}}\right)^{-\alpha_1}
        \left({m_1\over m_0}\right)^{-\alpha_2}
        \left({m_2\over m_1}\right)^{-\alpha_3}\right] 
        \left({m\over m_2}\right)^{-\alpha_4} 
        &m_2 < m \le m_u,\\
          \end{array}\right.
\label{eq:imf_mult}
\end{equation}
where $k$ contains the desired scaling, and $dN=\xi(m)\,dm$ is the
number of stars in the mass interval $m$ to
$m+dm$. Eq.~\ref{eq:imf_mult} is the general form of a five-part
power-law form, but at present observations only support a three-part
power-law IMF (Kroupa 2002) with $m_l=0.01\,M_\odot, m_{\rm
H}=0.08\,M_\odot, m_0=0.5\,M_\odot$, and $\alpha_2=\alpha_3=\alpha_4$,
\begin{equation}
          \begin{array}{l@{\quad\quad,\quad}l}
\alpha_0 = +0.3\pm0.7   &0.01 \le m/M_\odot < 0.08, \\
\alpha_1 = +1.3\pm0.5   &0.08 \le m/M_\odot < 0.50, \\
\alpha_2 = +2.3\pm0.3   &0.50 \le m/M_\odot. \\
          \end{array}
\label{eq:imf}
\end{equation}
The multi-part power-law form is convenient because it allows an
analytic mass-generation function to be used which leads to very
efficient generation of masses from an ensemble of random
deviates. The multi-part power-law form also has the significant
advantage that various parts of the IMF can be changed without
affecting other parts, such as changing the number of massive stars by
varying $\alpha_4$ without affecting the form of the luminosity
function of low-mass stars. Other functional descriptions of the IMF
are in use (e.g. Chabrier 2001).

The initial period distribution function needs to be consistent with
the TA data, and a convenient form which also has an analytic
period-generation function is derived in K2,
\begin{equation}
f_{\rm P,birth} = 2.5 \, { \left(lP - 1 \right)
  \over 45 + \left(lP - 1 \right)^2},
\label{eq:fp}
\end{equation}
where $f_{\rm P,birth}\,dlP$ is the proportion of binaries among all
systems with periods in the range $lP$ to $lP+dlP$ ($P$ in days
throughout this paper), and $1\le lP\equiv {\rm log}_{10}P$. The usual
notation for the binary proportion is used here, $f_{\rm P}=N_{\rm
bin, P}/N_{\rm sys}$, where $N_{\rm sys}=N_{\rm bin}+N_{\rm sing}$ is
the number of systems and $N_{\rm bin, P}$ is the number of binary
systems with periods in the bin $lP$.  The condition $\int_{lP}f_{\rm
P,birth}\,dlP=1$ (all stars being born in binaries) gives $P_{\rm
max}=10^{8.43}$~d for the maximum period obtained from the
distribution~\ref{eq:fp}.  $N-$body experiments demonstrate that the
observed range of periods ($P\approx 10^{0-9}$~d) must be present as a
result of the star-formation process; encounters in very dense
sub-groups cannot sufficiently widen initially more restricted period
distributions and at the same time lead to the observed fraction of
binaries in the Galactic field (Kroupa \& Burkert 2001).  Observations
show that the eccentricity distribution of Galactic-field binary
systems is approximately thermal, $f_{\rm e} = 2\,e$, and $N-$body
calculations demonstrate that such a distribution must be primordial
because encounters of young binaries in their embedded clusters cannot
thermalize an initially different distribution (K2; Kroupa \& Burkert
2001).

Binary systems in the Galactic field with short periods ($P\simless
10^3$~d) do show departures from simple pairing by having a
bell-shaped eccentricity distribution and a mass-ratio distribution
that appears to deviate from random sampling from the IMF. This is
apparent most dramatically in the eccentricity--period diagram which
shows an upper eccentricity-envelope for short-period binaries
(Duquennoy \& Mayor 1991).  This indicates that
binary-system--internal processes may have evolved a primordial
distribution. Such processes are envelope--envelope or disk--disk
interactions during youth, shared accretion during youth, rapid tidal
circularisation during youth, and slow tidal circularisation during
the main-sequence phase. These system-internal processes that change
the orbital parameters cannot be expressed with only a few equations
given the extremely complex physics involved, but a simple analytical
description is available through the K2-formulation of {\it
eigenevolution--feeding}. Feeding allows the mass of the secondary to
grow, while eigenevolution allows the eccentricity to circularise and
the period to decrease at small peri-astron distances, and merging to
occur if the semi-major axis of the orbit is smaller than~10 Solar
radii. About 3~per cent of initial binaries merge to form a single
star.  The eigenevolved model-main-sequence eccentricity--period
diagram, and the eccentricity and mass-ratio distributions of
short-period systems, agree well with observational data. In
particular, although the minimum period obtained from eq.~\ref{eq:fp}
is $P=10$~d, eigenevolution leads to the correct number of $P<10$~d
periods.  The resulting IMF of all stars shows slight departures from
the input IMF (eq.~\ref{eq:imf}) as a result of the mass-growth
(feeding) of some secondaries, but the deviations are well within the
IMF uncertainties.

%%%%%%%%%%%%%%%%%%%%%%%%%%%%%%%%%%%%%%%%%%%%%%%%%%%%%%%%%%%%%%%%%%%%%%%%%%%
\section{Models of Taurus--Auriga-like aggregates}
\label{sec:mods}

A star-formation event produces a spatial distribution of stars which
is typically centrally concentrated. As a first approach, this density
distribution is assumed to be smooth without sub-structure, because
this minimises the number of parameters needed.  Each binary system is
thus provided with phase-space variables which define its initial
location and velocity relative to the other binaries. A mathematically
very convenient smooth spatial distribution function is given by the
Plummer model,
\begin{equation}
\rho(R) = {3\,N_{\rm bin}\over 4\,\pi\,R_{\rm pl}^3} 
          {1\over [1 + (R/R_{\rm pl})^2]^{5/2}},
\label{eq:pldens}
\end{equation}
where $N_{\rm bin}$ is the number of binary systems in the aggregate
or cluster, and $R_{0.5} = 1.305\,R_{\rm pl}$ is the half-mass radius.
For each position the associated velocity needs to satisfy Poisson's
equation; the generating functions for position and velocity variables
are presented in Aarseth, H\'enon \& Wielen (1974).

To study the dynamical evolution of TA-like aggregates 25~binaries are
distributed as a Plummer density distribution embedded in an
analytical background gas potential with the same $R_{0.5}$ as the
stars, but twice the stellar mass.  The star-formation efficiency in
the aggregate is thus assumed to be $33$~per cent ($\epsilon=0.33$;
Lada 1999).  The velocity of each binary-star centre-of-mass is
$v_{\rm bin, g} = v_{\rm bin}/\sqrt{\epsilon}$ to insure initial
dynamical equilibrium, where $v_{\rm bin}$ is the velocity in an
equilibrium aggregate without gas.  Assuming dynamical equilibrium is
equivalent to assuming that the stars form from condensations in a
cloud core in virial equilibrium.  The gas potential is kept constant
for a time $\tau_{\rm D}$ after which it begins to evolve on an
exponential time-scale $\tau_{\rm M}$. Different values for $\tau_{\rm
D}$ and $\tau_{\rm M}$ test the effect on the binary population of
having the aggregates embedded in the potential for time scales
ranging from a fraction to about two crossing times. Each model is
integrated for 40~Myr, by which time the aggregates have largely
dissolved. All stars and binaries are kept in memory for data
reduction, for which a separate software package has been
written. Briefly, {\sc GasEx} writes out snapshots of all data at
regular time intervals. These are used to find all bound binary
systems and single stars and other data relevant for the present
discussion. We note that a more self-consistent treatment of
gas-expulsion using the smooth-particle-hydrodynamics approximation to
model the gas dynamics leads to the same results as when applying a
time-varying analytical background potential (Geyer \& Burkert 2001).

The models are listed in Table~\ref{tab:mods}. The half-mass radii and
number of systems are chosen to reflect the observed values (Gomez et
al. 1993; Hartmann, Ballesteros-Paredes \& Bergin 2001; Briceno et
al. 2002; Palla \& Stahler 2002). The durations of the embedded phase
($\approx t_{\rm D}+2\,\tau_{\rm M}$) approximately span the duration
of star formation and is typically longer than the crossing time --
the residual gas is removed over time through the accumulating stellar
outflows (Matzner \& McKee 2000).  There are 140~renditions per model
with different initial random number seeds.  The IMF power-law index
in the BD mass range, $\alpha_0$, is changed between some of the
models to allow an investigation of various BD contents.  The table
also contains the two KAH models of dense and rich ONC-type
clusters. These models are constructed in exactly the same way as the
present TA-like models, and thus allow a consistent comparison of the
evolution of the binary population and the apparent IMF with empirical
data to test for possible deviations from the standard model.

%======================== TABLE ==========================================
\begin{table*}
\hspace{-2.5cm}  
\begin{minipage}[h!]{15cm} \begin{center} \caption{Initial cluster
  models. For each except models A \& B there are 140 realisations
  with different initial random number seeds.  $N_{\rm sing}, N_{\rm
  bin}$ are the average number of single stars and binaries,
  respectively (each model shows deviations as a result of initial
  disruption due to overlapping binaries and binary-star {\it
  eigenevolution} during the pre-main sequence phase which can lead to
  a binary merging to a single star); the average stellar mass is
  $<\!\!m\!\!>$ -- for models T0 \& T1 the range of masses allowed is
  $0.01-5.0\,M_\odot$, while for models T2--T5 it is
  $0.01-2.0\,M_\odot$, and for the KAH models A \& B the range is
  $0.01-50\,M_\odot$; $\sigma_{\rm 3D}$ is the 3D velocity dispersion
  of systems; $t_{\rm cross}=2\,R_{0.5}/\sigma_{3D}$ is the nominal
  crossing time; $\rho_{\rm C}$: central number density; $M_{\rm st},
  M_{\rm g}$: mass in stars and gas, respectively, these are average
  values of the 140~renditions per model; $\tau_{\rm M}$: time-scale
  for gas expulsion (eq.~2 in KAH); $t_{\rm D}$: onset of
  gas-expulsion; the stellar and gas distributions have half-mass
  radii $R_{0.5}, R_{0.5, {\rm g}} (R_{\rm pl,g}(0)=0.766\,R_{0.5,{\rm
  g}})$, respectively. The BD population is described by the IMF
  power-law index $\alpha_0$, and the fraction of the population that
  are BDs is given by the last column. }

    \vspace*{0.2cm}
    \label{tab:mods}
    \begin{tabular}[h!]{ccccccccccccccc} \hline

model &$N_{\rm sing}$ &$N_{\rm bin}$ &$R_{0.5}$ &$<\!\!m\!\!>$ 
&$\sigma_{\rm 3D}$ &$t_{\rm cross}$ &log$_{10}\rho_{\rm C}$
&$M_{\rm st}$ &$M_{\rm g}$ &$R_{\rm 0.5,{\rm g}}$ 
&$\tau_{\rm M}$ &$t_{\rm D}$ &$\alpha_0$ &$N_{\rm BD}/N_{\rm st}$\\ \hline

      &    &              &[pc]      &[$M_\odot$]           
&[km/s]            &[Myr] &[stars/pc$^3$] 
&[$M_\odot$] &[$M_\odot$] &[pc] 
&[Myr] &[Myr] & &[per cent]\\ \hline

T0   &2   &23   &0.30    &0.32 &0.51  &1.2 &3.0
&16 &32 &0.30 &1.0 &0.5 &$+0.3$ &37\\

T1   &1   &24   &0.80    &0.32 &0.31  &5.2 &1.7 
&16 &32 &0.80 &1.0 &0.5 &$+0.3$ &37\\

T2   &1   &24   &0.30    &0.35 &0.53  &1.1 &3.0
&18 &36 &0.30 &2.0 &2.0 &$-4.2$ &10\\

T3   &1   &24   &0.30    &0.35 &0.53  &1.1 &3.0
&18 &36 &0.30 &2.0 &2.0 &$-3.0$ &12\\

T4   &1   &24   &0.30    &0.33 &0.51  &1.2 &3.0
&17 &34 &0.30 &2.0 &2.0 &$-1.5$ &18\\

T5   &1   &24   &0.30    &0.30 &0.49  &1.2 &3.0
&15 &30 &0.30 &2.0 &2.0 &$-0.5$ &26\\

\hline

A   &575   &4642   &0.450    &0.38 &6.8  &0.13 &4.8 
&3746 &7492 &0.450 &0.045 &0.60  &$+0.3$ &37\\

B   &1247  &4298   &0.206    &0.42 &10.8 &0.038 &5.8
&4170 &8340 &0.206  &0.021 &0.60 &$+0.3$ &37\\

    \end{tabular}
  \end{center}
  \end{minipage}
\end{table*}
%=========================================================================

%%%%%%%%%%%%%%%%%%%%%%%%%%%%%%%%%%%%%%%%%%%%%%%%%%%%%%%%%%%%%%%%%%%%%%%%%%%
\section{Dynamical evolution}
\label{sec:ev}

An outline of the dynamical evolution and associated changes of the
stellar population is given in this section.

\begin{figure}
\begin{center}
\rotatebox{0}{\resizebox{0.6 \textwidth}{!}
{\includegraphics{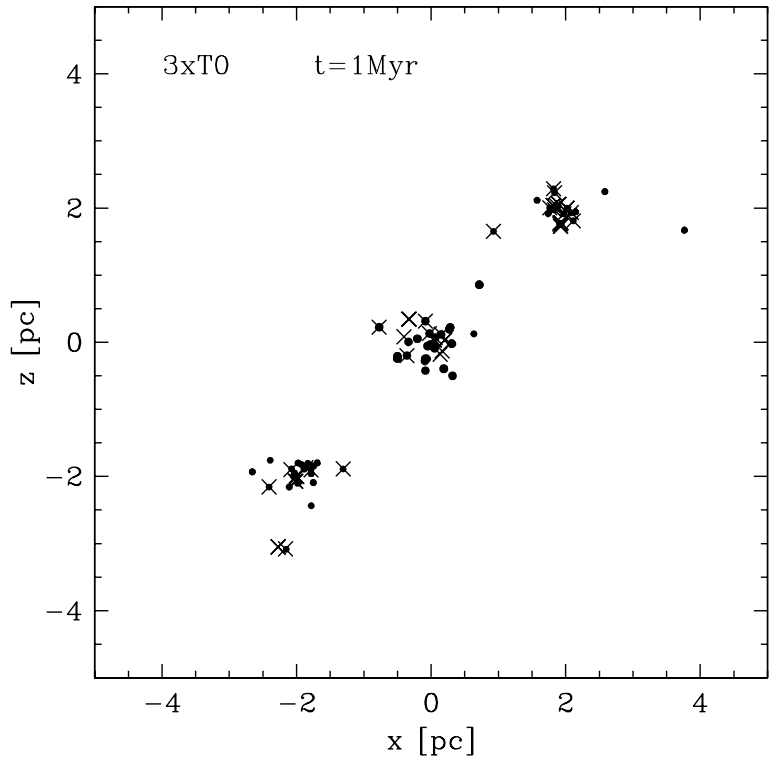}}}
\vskip -20mm
\caption
{The appearance of the first three of the 140 T0~aggregates after
1~Myr.  Crosses denote BDs, many of which are in binaries with stellar
(dots) primaries.  At this age each aggregate contains about
$20\,M_\odot$ of gas and is thus still embedded, with the fraction of
gas mass being slightly larger than the mass in stars and BDs
($16\,M_\odot$).  The relative motion of the aggregates is
neglected. The $z-$direction is perpendicular to the Galactic plane,
while the $x-$direction is arbitrary within the Galactic plane. }
\label{fig:app1}
\end{center}
\end{figure}

\begin{figure}
\begin{center}
\rotatebox{0}{\resizebox{0.6 \textwidth}{!}
{\includegraphics{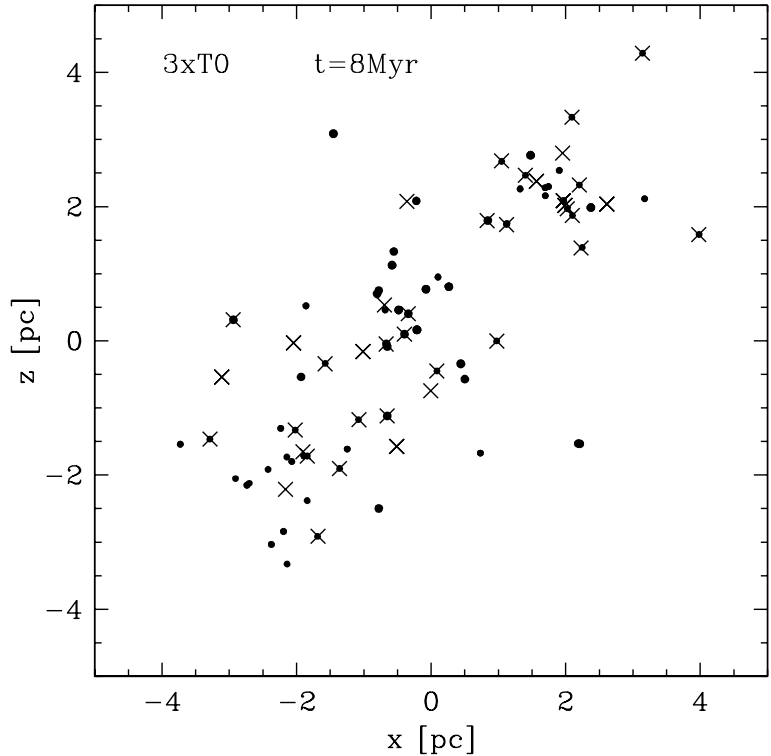}}}
\vskip -20mm
\caption
{As fig.~\ref{fig:app1} but at 8~Myr.}
\label{fig:app2}
\end{center}
\end{figure}

\begin{figure}
\begin{center}
\rotatebox{0}{\resizebox{0.5 \textwidth}{!}
{\includegraphics{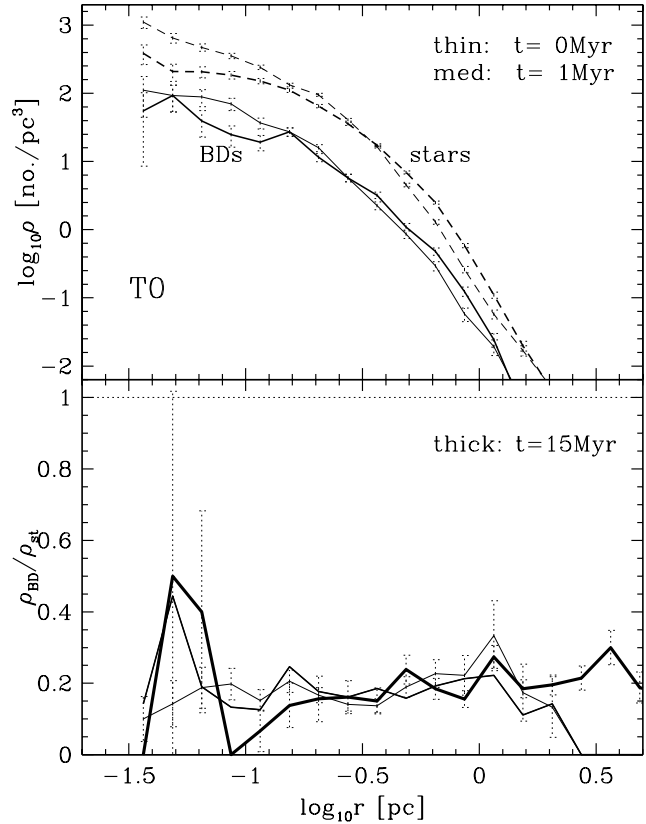}}}
\vskip 0mm
\caption
{{\bf Upper panel:} The radial density profile of stellar ($0.08 \le
m_p/M_\odot \le 1.0$, dashed lines) and BD systems ($0.01 \le
m_p/M_\odot \le 0.08$, solid lines). A system consists either of a
single star or BD or a binary with a stellar or BD primary mass. {\bf
Lower panel:} The ratio of the density of stellar- and BD-systems.  In
both panels the curves are averages of 140~different random-number
renditions of model~T0.}
\label{fig:rprT0}
\end{center}
\end{figure}

\begin{figure}
\begin{center}
\rotatebox{0}{\resizebox{0.5 \textwidth}{!}
{\includegraphics{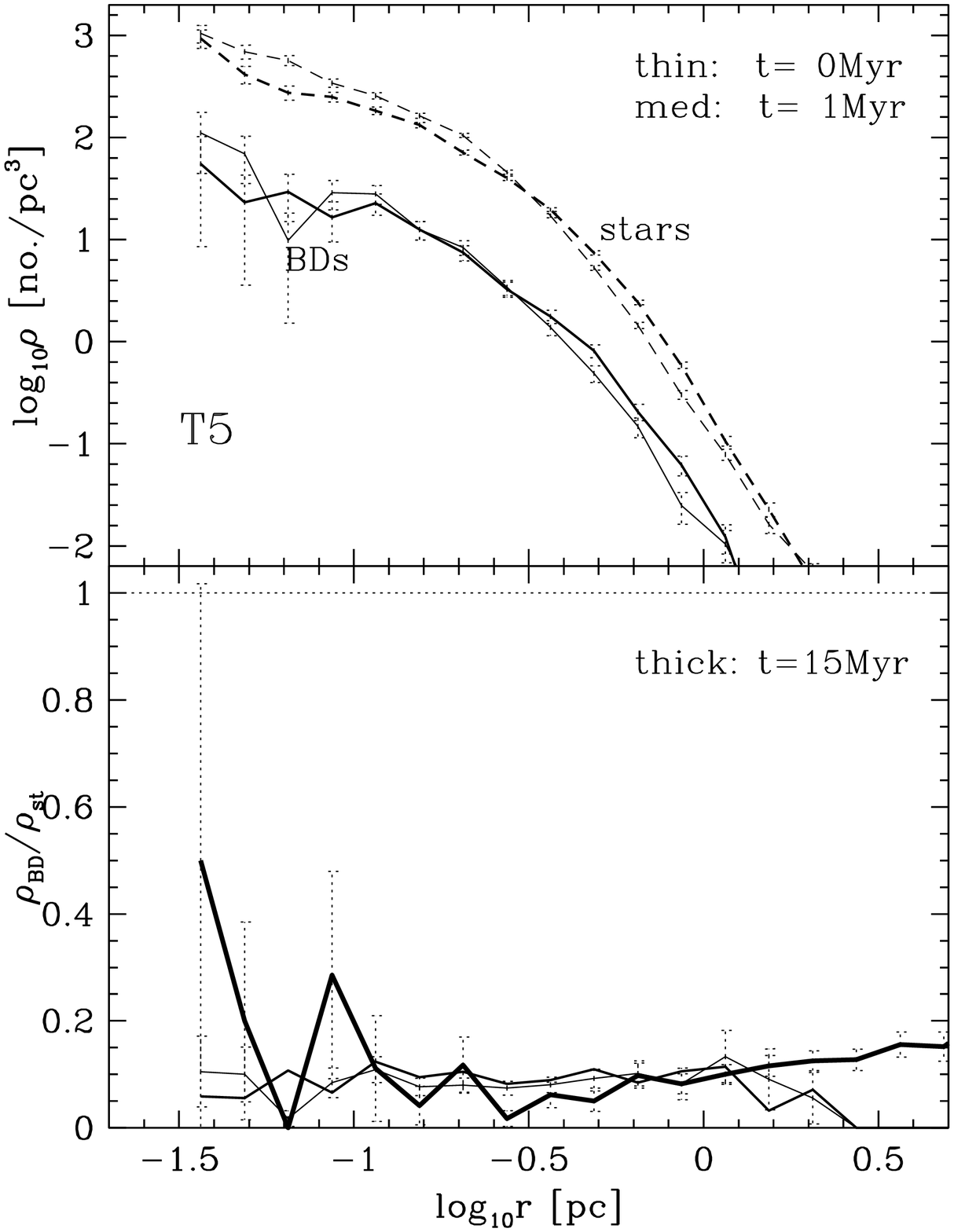}}}
\vskip 0mm
\caption
{As Fig.~\ref{fig:rprT0} but for model~T5. 
}
\label{fig:rprT5}
\end{center}
\end{figure}

The aggregates evolve by expanding due to two-body encounters and the
removal of the gas. Snapshots of this are shown in Figs~\ref{fig:app1}
and~\ref{fig:app2}. The density profiles of stellar- and BD-systems
are plotted at various times in Figs.~\ref{fig:rprT0} (model~T0)
and~\ref{fig:rprT5} (model~T5). The dissolution of the aggregates is
evident by the reduction with time of the number densities, and the
figure also shows that the BD and stellar populations do not separate
significantly in these models.  That is, the BDs trace the stars,
except for a small high-velocity BD distribution which results from
BDs being ejected in binary--binary encounters (Figs.~\ref{fig:fv1}
and~\ref{fig:fv2} below).
%=================================== FIGURE ==============================
% /NB_EVAL/Bouv/Taurus/plot_sing1Myr.sm
%=========================================================================
%=================================== FIGURE ==============================
% /NB_EVAL/Bouv/Taurus/plot_sing8Myr.sm
%=========================================================================
%=================================== FIGURE ==============================
% /NB_EVAL/Bouv/Taurus/radprofT0.sm
%=========================================================================
%=================================== FIGURE ==============================
% /NB_EVAL/Bouv/Taurus/radprofT5.sm
%=========================================================================

\begin{figure}
\begin{center}
\rotatebox{0}{\resizebox{0.5 \textwidth}{!}
{\includegraphics{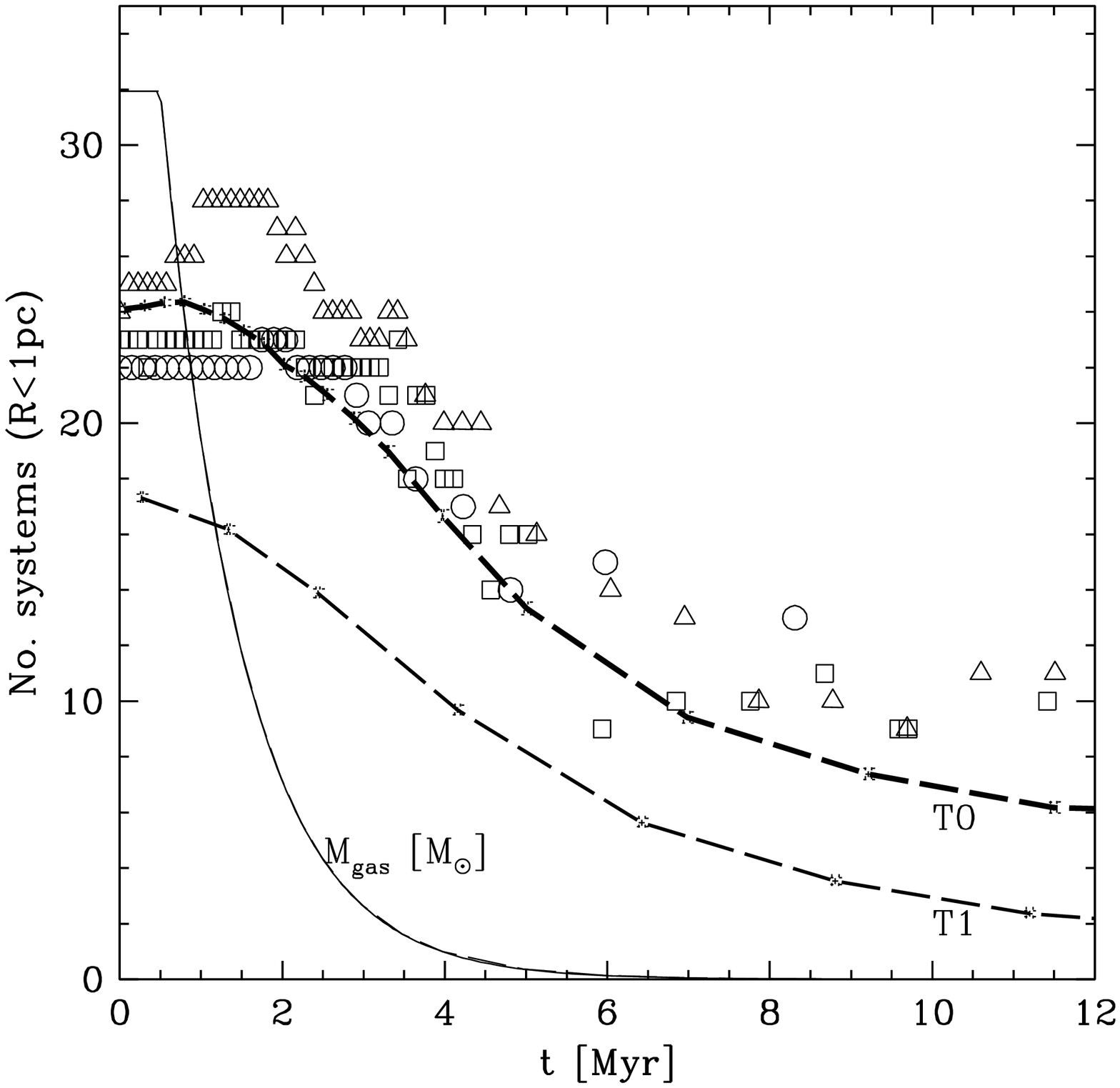}}}
\vskip -20mm
\caption
{The number of systems (all masses) located within 1~pc of the density
centre of the ensemble of TA-like aggregates is shown as the dashed
lines. The thick curve denotes the average of 140~T0 models, while the
thin curve is the average of 140~T1 models. Errorbars are standard
deviation of the mean values.  Open triangles, squares and circles are
the three T0~models shown in Figs.~\ref{fig:app1} and~\ref{fig:app2}.
The gas mass of one of the models~T0 is shown as the thin solid
curve.}
\label{fig:nsys1}
\end{center}
\end{figure}

\begin{figure}
\begin{center}
\rotatebox{0}{\resizebox{0.5 \textwidth}{!}
{\includegraphics{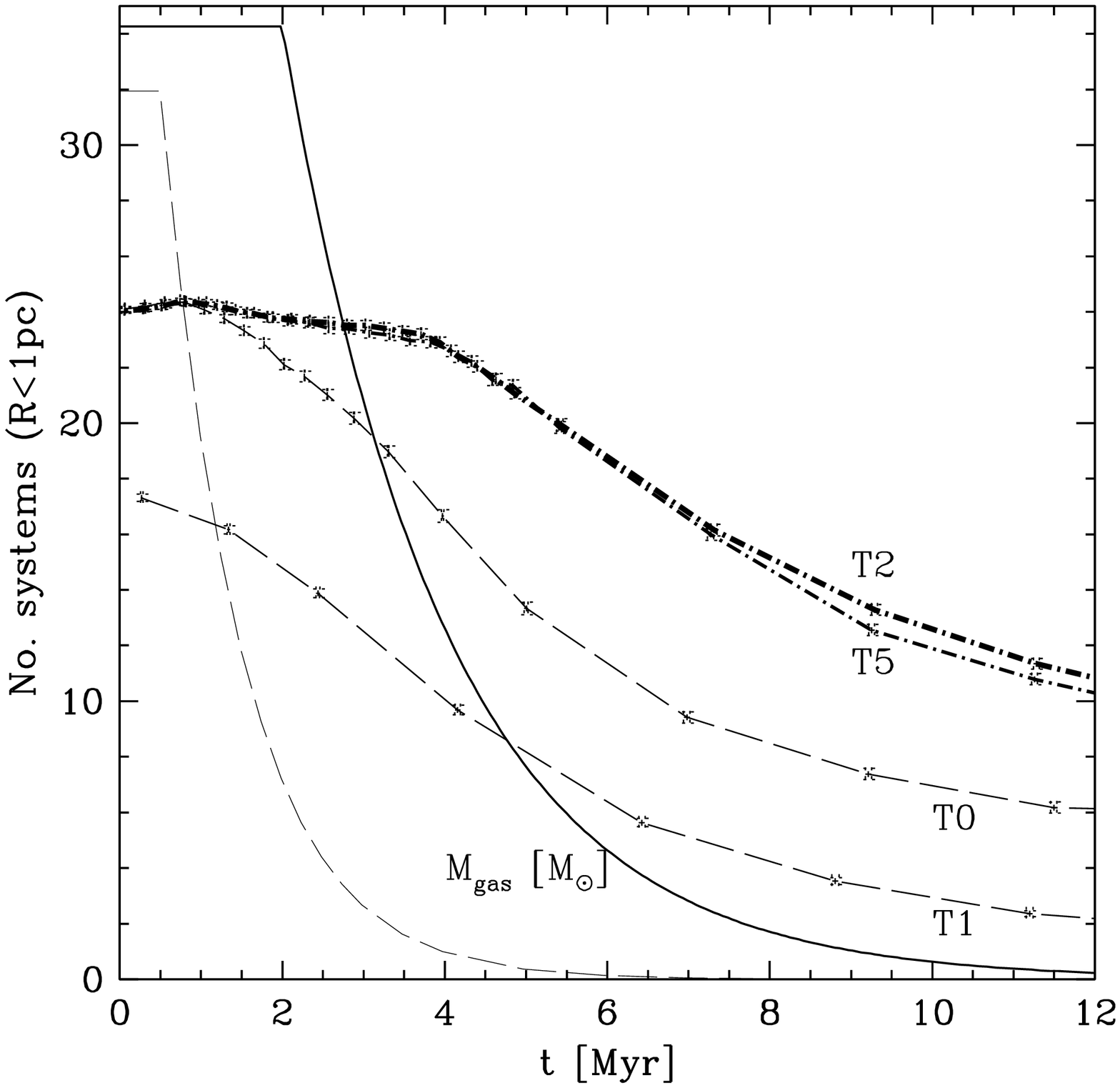}}}
\vskip -20mm
\caption
{As Fig.~\ref{fig:nsys1} but showing models~T2 and~T5 in comparison
with models~T0 and~T1. Note the longer embedded phase and the
resulting longer life-time of~T2 and~T5. The gas mass of one of the
140~T2 models is shown as the solid curve and is initially
$34\,M_\odot$, while the dashed curve is $M_{\rm gas}(t)$ for one of
the T0~models.}
\label{fig:nsys2}
\end{center}
\end{figure}

The nominal dissolution time due to redistribution of energy within
the aggregates due to two-body encounters is $t_{\rm
diss}\approx100\,t_{\rm relax}$, where $t_{\rm relax} \approx 0.1
N\,t_{\rm cross}/{\rm ln}N$ is the median two-body relaxation time
(Binney \& Tremaine 1987). For models T0-T5 $t_{\rm relax}\approx
0.8-1.3$~Myr for $N_{\rm sys}=25-50$, so that $t_{\rm diss}\approx
80-130$~Myr. Gas removal unbinds the aggregates on a much shorter
time-scale though, as is shown in Figs~\ref{fig:nsys1}
and~\ref{fig:nsys2}. Thus, by 32~Myr models~T0, T1, T2 and T5 contain
on average only about~2,~5,~7 and~7 systems within the central 1~pc
sphere, respectively.
%=================================== FIGURE ==============================
% /NB_EVAL/Bouv/Taurus/pl_diagn1_nsys.sm
%=========================================================================
%=================================== FIGURE ==============================
% /NB_EVAL/Bouv/Taurus/pl_diagn1_nsysT2T5.sm
%=========================================================================

\begin{figure}
\begin{center}
\rotatebox{0}{\resizebox{0.5 \textwidth}{!}
{\includegraphics{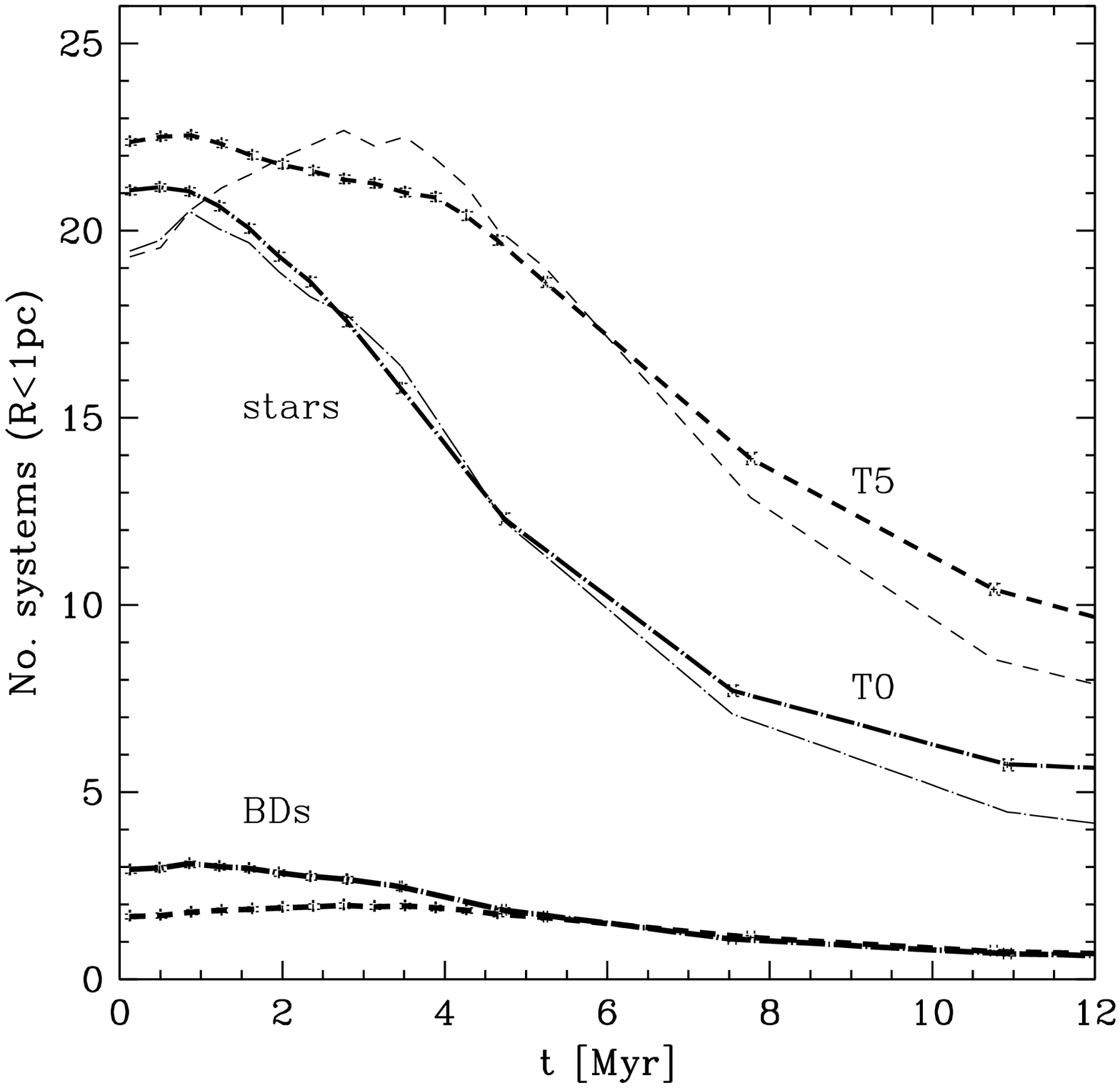}}}
\vskip -20mm
\caption
{The number of BD systems (primaries with masses $0.02-0.08\,M_\odot$)
within a distance of 1~pc of the aggregate centres in models~T0 and~T5
are plotted as the thick curves (dot-dashed: T0; dashed: T5). They are
(arbitrarily) scaled to become the corresponding thin curves to show
the evolution relative to the number of stellar systems (primaries
with masses $0.08-10\,M_\odot$) depicted as the upper thick curves.  }
\label{fig:nsysBDs}
\end{center}
\end{figure}

The number of BD systems within the central 1~pc sphere increases at
first as a result of BD companions being freed from binary systems,
but at later times ($t\simgreat 5$~Myr) the relative number of BDs
decays more rapidly than the number of stellar systems as a result of
energy equipartition in the aggregates and gas removal
(Fig.~\ref{fig:nsysBDs}). During the embedded phase the BDs acquire on
average somewhat higher velocities than the stars leading to their
slightly more rapid dispersal after the gas has been removed
(\S~\ref{sec:distr}).  The net effect is nevertheless such that at 0,
1 and about 15~Myr the BD- and stellar-systems have statistically
indistinguishable density profiles (Figs~\ref{fig:rprT0}
and~\ref{fig:rprT5}).
%=================================== FIGURE ==============================
% /NB_EVAL/Bouv/Taurus/pl_diagn17_nsysT0T5.sm
%=========================================================================

\begin{figure}
\begin{center}
\rotatebox{0}{\resizebox{0.5 \textwidth}{!}
{\includegraphics{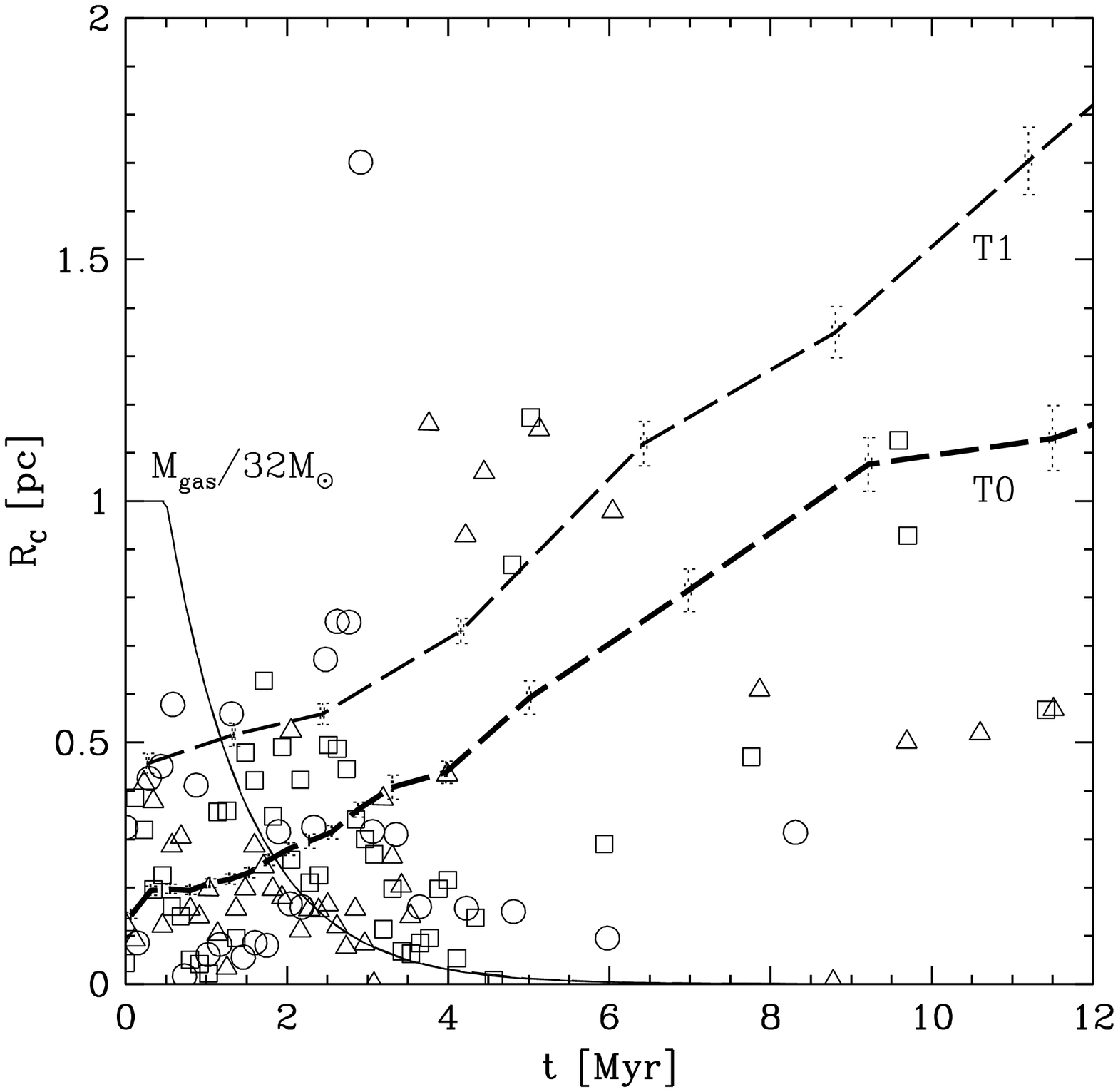}}}
\vskip -20mm
\caption
{The evolution of the core radius.  The thick dashed curve shows the
average of 140~T0 models, while the thin dashed curve is the average
of 140~T1 models. Errorbars are standard deviation of the mean values.
Open triangles, squares and circles are the three T0~models shown in
Figs.~\ref{fig:app1} and~\ref{fig:app2}.  The gas mass of one of the
models~T0 is shown as the thin solid curve.}
\label{fig:rc1}
\end{center}
\end{figure}

\begin{figure}
\begin{center}
\rotatebox{0}{\resizebox{0.5 \textwidth}{!}
{\includegraphics{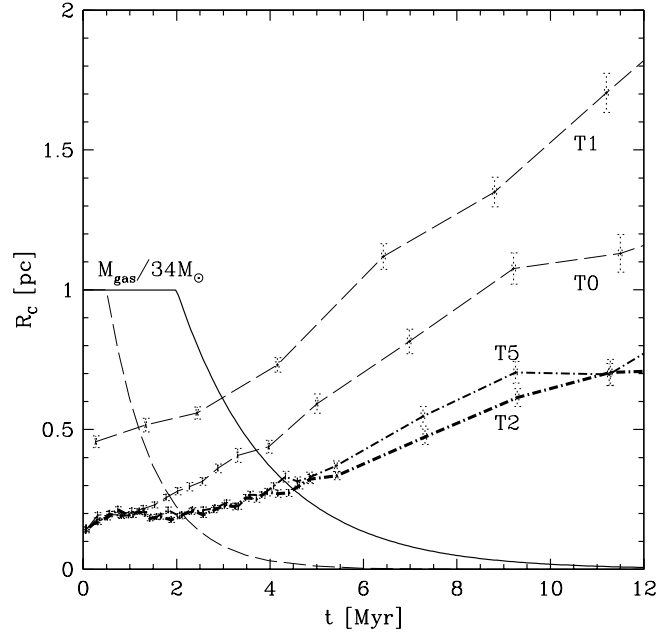}}}
\vskip -20mm
\caption
{As Fig.~\ref{fig:rc1} but showing models~T2 and~T5 in comparison with
models~T0 and~T1. The gas mass of one of the 140~T2 models is shown as
the solid curve and is initially $34\,M_\odot$, while the dashed curve
plots $M_{\rm gas}(t)$ for one of the T0~models.}
\label{fig:rc2}
\end{center}
\end{figure}

The core radius (eq.~11 in KAH), which indicates that radius by which
the luminosity density has fallen to half its central value, is shown
in Fig.~\ref{fig:rc1}. It expands substantially and becomes larger
than 1~pc by 6--8~Myrs for models~T0 and~T1.  The slowed expansion due
to delayed gas-expulsion in models~T2--T5 is evident in
Fig.~\ref{fig:rc2}, these models retain a more compact morphology and
the core radius reaches 1~pc by about 20~Myr.  

\begin{figure}
\begin{center}
\rotatebox{0}{\resizebox{0.5 \textwidth}{!}
{\includegraphics{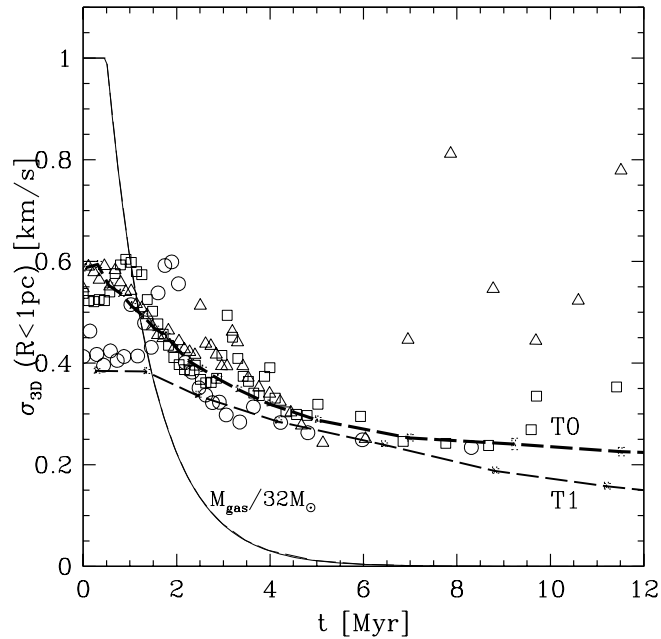}}}
\vskip -20mm
\caption
{The three-dimensional velocity dispersion of systems located within
1~pc of the density centre of each TA-like aggregate is shown as the
dashed lines. The thick curve denotes the average of 140~T0 models,
while the thin curve is the average of 140~T1 models. Errorbars are
standard deviation of the mean values.  Open triangles, squares and
circles are the three T0~models shown in Figs.~\ref{fig:app1}
and~\ref{fig:app2}.  The gas mass of one of the models~T0 is shown as
the thin solid curve.}
\label{fig:vd1}
\end{center}
\end{figure}

\begin{figure}
\begin{center}
\rotatebox{0}{\resizebox{0.5 \textwidth}{!}
{\includegraphics{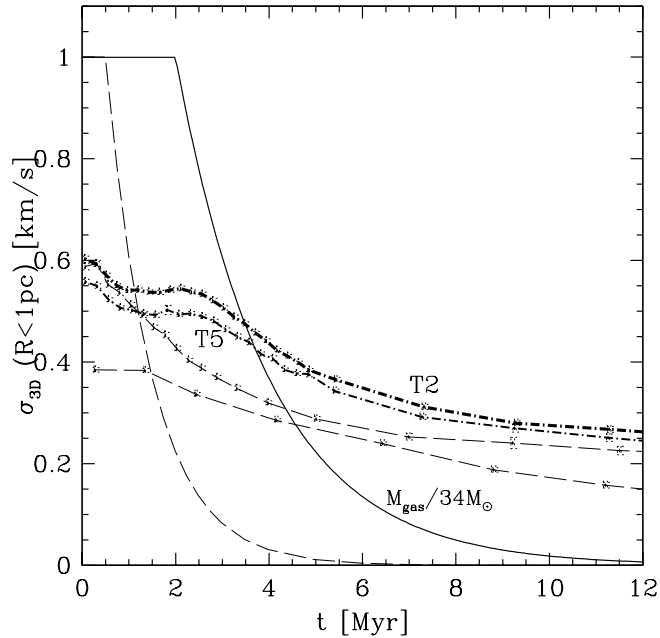}}}
\vskip -20mm
\caption
{As Fig.~\ref{fig:vd1} but showing models~T2 and~T5 in comparison with
models~T0 and~T1. Note the longer embedded phase and the resulting
larger velocity dispersion of~T2 and~T5. The gas mass of one of the
140~T2 models is plotted as the solid curve, while the dashed curve
depicts the gas mass for one of the T0~models. }
\label{fig:vd2}
\end{center}
\end{figure}

The evolution of the three-dimensional velocity dispersion within the
central 1~pc sphere is shown in Figs.~\ref{fig:vd1}
and~\ref{fig:vd2}. It drops from an initial 0.6~km/s (models~T0,
T2--T5) or 0.4~km/s (model~T1) to values close to 0.3~km/s or less.
These values are consistent with the low stellar velocity dispersion
of 0.2~km/s (one dimensional) suggested within the TA cloud filaments
by Hartmann (2002).

\begin{figure}
\begin{center}
\rotatebox{0}{\resizebox{0.5 \textwidth}{!}
{\includegraphics{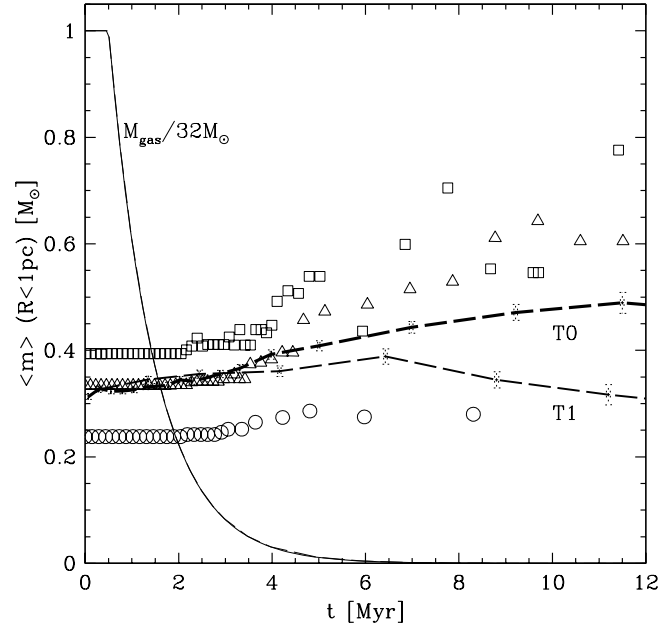}}}
\vskip -20mm
\caption
{The mean mass of systems located within 1~pc of the density centre of
each TA-like aggregate is shown as the dashed lines. Thick curves
denote the average of the 140~T0 models, while the thin curve is the
average of 140~T1 models. Errorbars are standard deviation of the mean
values.  Open triangles, squares and circles are the three T0~models
shown in Figs.~\ref{fig:app1} and~\ref{fig:app2}. Note the difference
in the initial values which stem from statistical variations and
feeding.  The gas mass of one of the models~T0 is shown as the thin
solid curve.}
\label{fig:mm1}
\end{center}
\end{figure}

\begin{figure}
\begin{center}
\rotatebox{0}{\resizebox{0.5 \textwidth}{!}
{\includegraphics{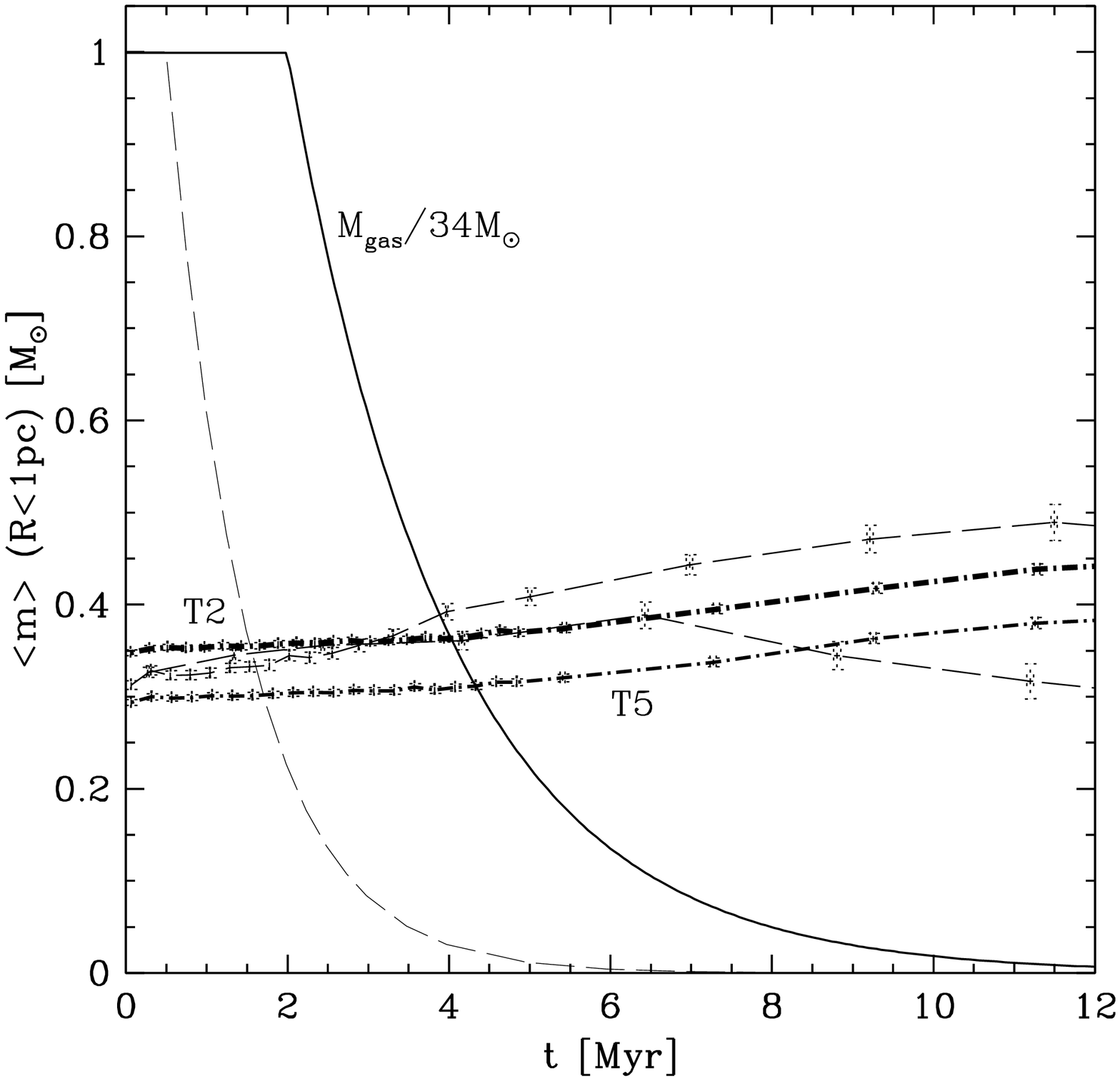}}}
\vskip -20mm
\caption
{As Fig.~\ref{fig:mm1}: models~T2 and~T5 in comparison with models~T0
and~T1. The gas mass of one of the 140~T2 models is depicted as the
solid curve and is initially $34\,M_\odot$. The dashed curve is the
gas mass for one of the T0~models.}
\label{fig:mm2}
\end{center}
\end{figure}

The mean system mass within the central 1~pc region indicates the
level to which mass segregation develops. It increases only by a small
amount in all models (Figs.~\ref{fig:mm1}, \ref{fig:mm2}).  By about
32~Myr models T0 and T2--T5 have a mean system mass of
$0.39-0.45\,M_\odot$, while the initially least concentrated model~T1
(which has the same IMF as~T0) has $<m>\approx0.25\,M_\odot$. Thus,
initially more compact aggregates produce final long-lived small$-N$
groups with more massive components, which is a result of the larger
probability of partner exchanges in the denser environment.
%=================================== FIGURE ==============================
% /NB_EVAL/Bouv/Taurus/pl_diagn14_rc.sm
%=========================================================================
%=================================== FIGURE ==============================
% /NB_EVAL/Bouv/Taurus/pl_diagn14_rcT2T5.sm
%=========================================================================
%=================================== FIGURE ==============================
% /NB_EVAL/Bouv/Taurus/pl_diagn2_vdisp.sm
%=========================================================================
%=================================== FIGURE ==============================
% /NB_EVAL/Bouv/Taurus/pl_diagn2_vdispT2T5.sm
%=========================================================================
%=================================== FIGURE ==============================
% /NB_EVAL/Bouv/Taurus/pl_diagn3_meanmass.sm
%=========================================================================
%=================================== FIGURE ==============================
% /NB_EVAL/Bouv/Taurus/pl_diagn3_meanmassT2T5.sm
%=========================================================================

%%%%%%%%%%%%%%%%%%%%%%%%%%%%%%%%%%%%%%%%%%%%%%%%%%%%%%%%%%%%%%%%%%%%%%%%%%%
\section{The binary systems}
\label{sec:bins}

\begin{figure}
\begin{center}
\rotatebox{0}{\resizebox{0.5 \textwidth}{!}
{\includegraphics{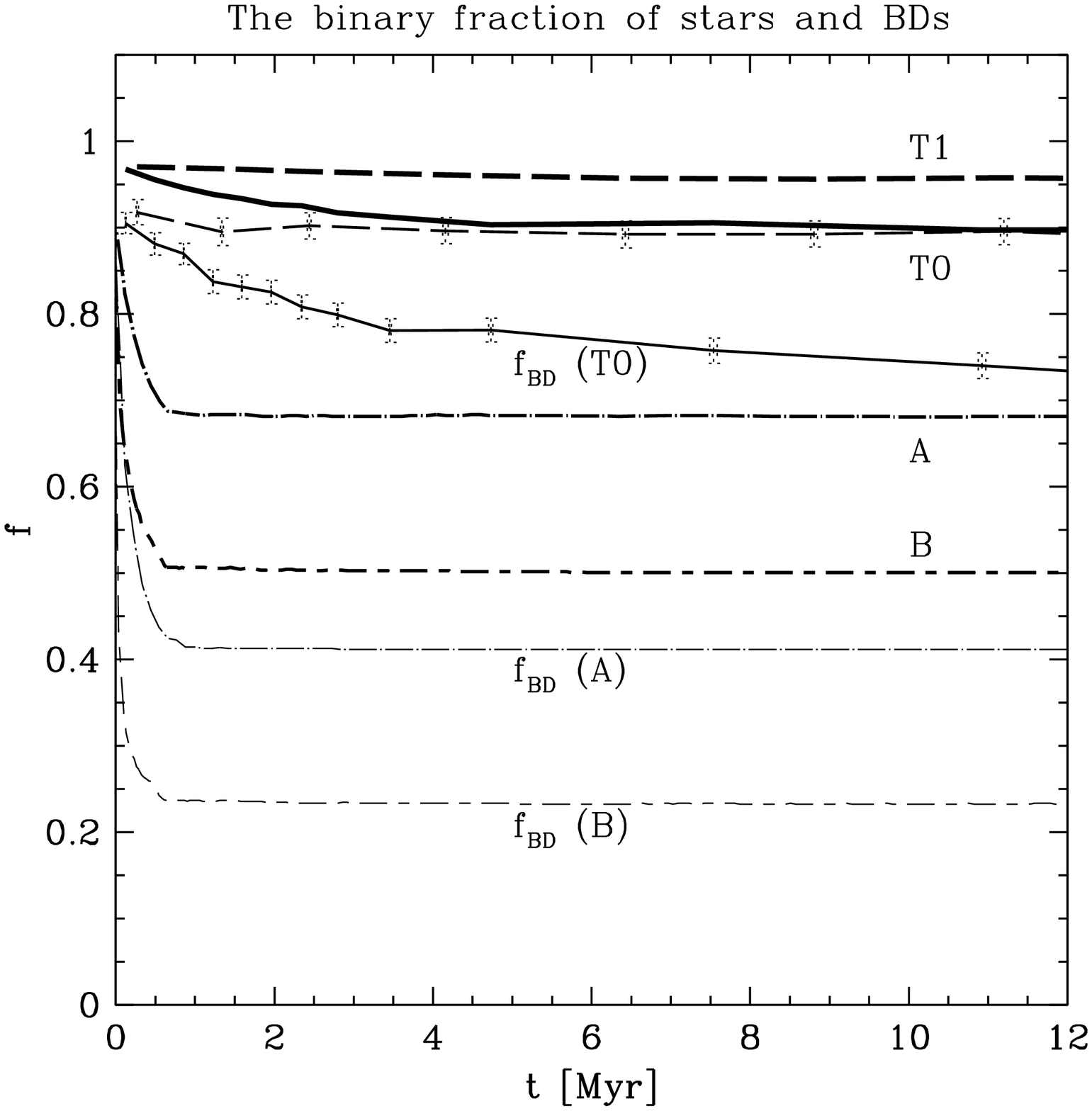}}}
\vskip -20mm
\caption
{The evolution of the binary fraction for models~T0 and~T1. Thick
curves are the binary fraction of stellar primaries ($0.15\le
m_1/M_\odot \le 1.0$), while thin curves are the binary fraction of BD
primaries, $f_{\rm BD}$ ($0.02 \le m_1/M_\odot \le 0.08$). The curves
are averages of 140~renditions of each model, and the errorbars are
standard deviations of the mean. For comparison, the evolution of the
corresponding quantities in rich ONC-like clusters is indicated by
models A and B from KAH.}
\label{fig:f1}
\end{center}
\end{figure}

\begin{figure}
\begin{center}
\rotatebox{0}{\resizebox{0.5 \textwidth}{!}
{\includegraphics{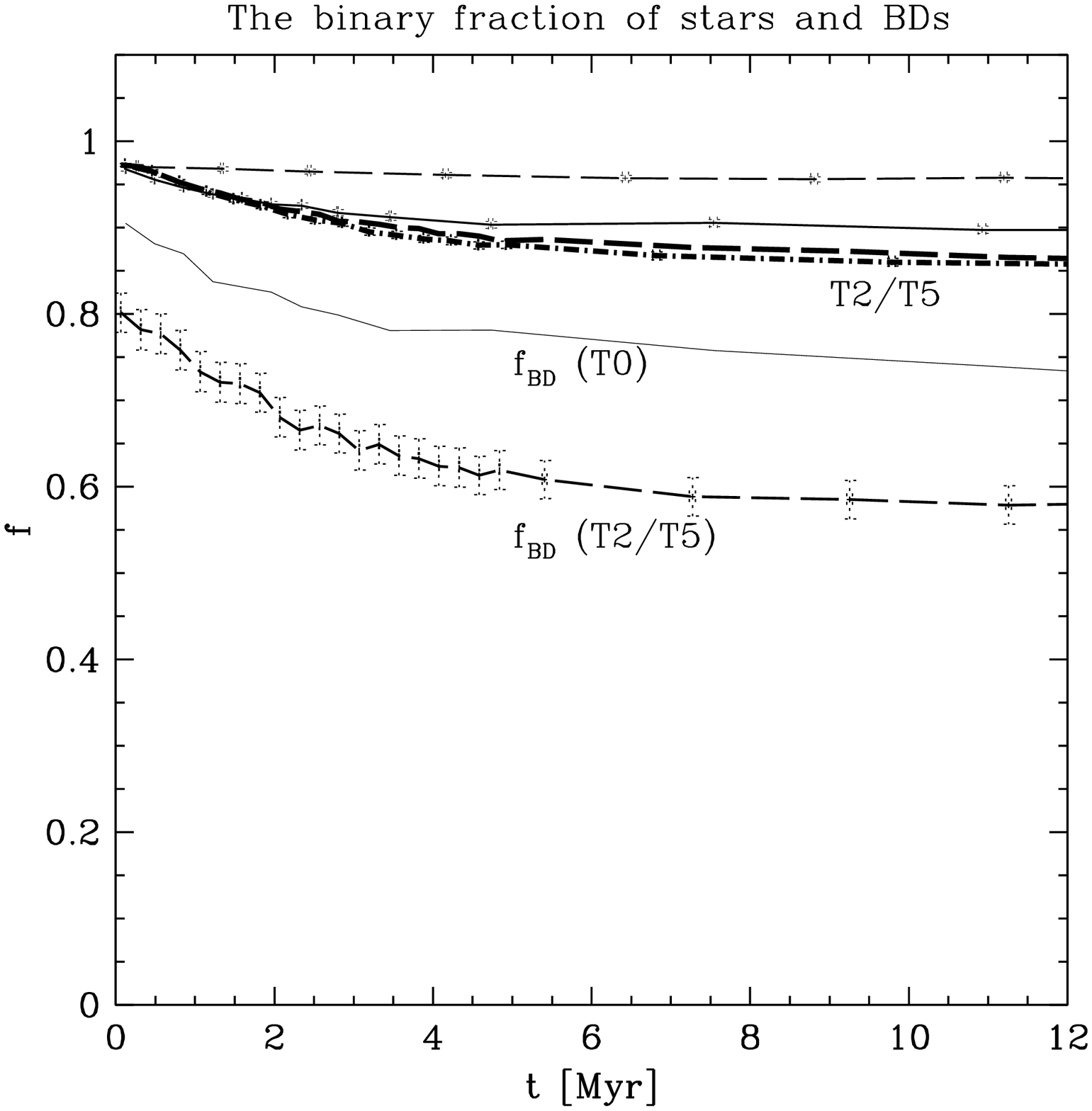}}}
\vskip -20mm
\caption
{As Fig.~\ref{fig:f1} but showing models~T2 and~T5 in comparison with
models~T0 and~T1 ($f_{\rm BD}(T1)$ is not plotted for clarity). }
\label{fig:f2}
\end{center}
\end{figure}

The binary-star population evolves through system-internal processes
(eigenevolution, \S~\ref{sec:standmod}). This affects only the
short-period binaries ($P\simless 10^3$~d). Encounters between
binaries disrupt long-period binaries on the crossing time-scale of
the aggregate.  Binary-system disruption extends approximately down to
a characteristic (or ``thermal'') period, $P_{\rm th}$ [d], at which
the circular orbital velocity, $v_{\rm orb}$ [km/s], of a binary
system with mass $m_{\rm sys}$ [$M_\odot$], equals the velocity
dispersion, $\sigma_{\rm 3D}$ [km/s], in the cluster. Because
$\sigma_{\rm 3D}$ decreases as the aggregate expands, $P_{\rm th}$
moves to longer periods until it extends beyond the cutoff period of
the evolving period distribution function, which is moving towards
$P_{\rm th}$ (Kroupa 2000).  There is not a sharp truncation of the
period distribution function however --- binaries with $P>P_{\rm th}$
located in the outer region of the cluster at $t=0$ may arrive in the
inner parts of the cluster when $P<P_{\rm th}$ making them resistant
against disruption.  When the crossover of $P_{\rm th}$ and $P_{\rm
cut}$ happens, further binary-disruption is mostly halted, and only
rare close encounters in the surviving stellar group lead to further
changes in the orbital parameters of a few binaries.  With
\begin{equation}
{\rm log}_{10}P = 6.986 + {\rm log}_{10}m_{\rm sys} - 3\,{\rm
log}_{10}v_{\rm orb},
\label{eq:vorb}
\end{equation} 
$P=P_{\rm th}$ if $v_{\rm orb}=\sigma_{\rm 3D}$. For $\sigma_{\rm
3D}=0.6$~km/s and $m_{\rm sys}=0.4\,M_\odot$ (Table~\ref{tab:mods})
$P_{\rm th}=10^{7.3}$~d, which means that a large number of binaries
with periods $7.3 \simless lP \simless 8.3$ will be disrupted implying
reduction of $f_{\rm P}$ by about 20~per cent. The final
binary-fraction should thus be $f\approx0.8$. This is confirmed by the
data plotted in Figs.~\ref{fig:f1} and~\ref{fig:f2}, which also
confirm that prior dynamical evolution of the present-day TA groups
did not lead to significant changes of the binary-star population.
The figures show that BDs retain a significantly smaller binary
proportion than stellar primaries. This is a result of the weaker
binding energy of BD--BD binaries. Fig.~\ref{fig:f2} also shows that
models~T2 and~T5 have a significantly smaller BD binary proportion
than models~T0 and~T1. This comes about because models T2--T5 enjoy a
longer embedded phase leading to more encounters per system. An effect
also playing a role is that T2--T5 have relatively fewer BDs by virtue
of the steeply declining IMF for $m<0.08\,M_\odot$. A larger fraction
of stars is thus obtained implying a larger average mass of the
siblings that a typical BD binary encounters, and thus larger relative
perturbative forces which are more efficient in destroying the BD
binaries than in models~T0 and~T1.

%=================================== FIGURE ==============================
% /NB_EVAL/Bouv/Taurus/pl_diagn19f.sm
%=========================================================================
%=================================== FIGURE ==============================
% /NB_EVAL/Bouv/Taurus/pl_diagn19fT2T5.sm
%=========================================================================
The evolution of the binary proportion of the ONC-like models of KAH
is also plotted in Fig.~\ref{fig:f1}. From Table~\ref{tab:mods},
$\sigma_{\rm 3D}=6.8$ (model~A) and~10.8 (model~B), implying $P_{\rm
th}=10^{4.1}$~d (model~A) and~$10^{3.5}$~d. Thus, in ONC-like clusters
the period distribution function is cut-off at a much shorter period
than in TA-like aggregates despite the same initial period
distribution.

To summarize, the important result here is that TA-like aggregates
could not have provided significantly to the Galactic-field population
which has a binary proportion of about $f=0.55-0.6$ (Duquennoy \&
Mayor 1991).

%%%%%%%%%%%%%%%%%%%%%%%%%%%%%%%%%%%%%%%%%%%%%%%%%%%%%%%%%%%%%%%%%%%%%%%%%%%
\section{The relative distribution of brown dwarfs and stars}
\label{sec:distr}

The relative distribution of stars and BDs is a potential discriminant
between different formation scenarios. For example, the {\it
embryo-ejection hypothesis} according to which BDs are ejected
unfinished stars from forming multiple systems (Reipurth 2000; Boss
2001; Reipurth \& Clarke 2001; Bate, Bonnell \& Bromm 2002) implies a
spatially wide distribution of BDs if their typical ejection
velocities are about 1~km/s or larger.

In contrast, the standard reference star-formation model
(\S~\ref{sec:standmod}), which is extended here and in KAH to include
BDs, assumes that stars and BDs form with the same kinematical and
binary properties at the point when the system can be treated as an
$N-$body system. For TA-like aggregates the standard model with BDs
implies that the BDs and stars trace approximately the same density
distribution (Figs.~\ref{fig:rprT0} and~\ref{fig:rprT5}).  Close
encounters between binaries eject, from the temporary four and later
three-body system, preferentially the least massive members,
i.e. typically the BDs (Sterzik \& Durisen 1998). The standard model
should thus yield, for the BDs, a high-velocity tail populated almost
exclusively by single BDs. This is evident in Figs.~\ref{fig:rprT0}
and~\ref{fig:rprT5} by the slightly increasing ratio $\rho_{\rm
BD}/\rho_{\rm st}$ with increasing $r$ at late times.  The ejected BDs
will also have truncated disks with a truncation radius of about the
peri-astron distance that produced the BD escaper. Thus, for an
ejection velocity $\ge1$~km/s the truncation radius will be $\le40$~AU
(eq.~\ref{eq:vorb} plus Kepler's third law).

This is basically the same result as for the embryo-ejection scenario
(Reipurth 2000) which was recently championed by Reipurth \& Clarke
(2001), making a distinction between both (standard model vs the
ejection hypothesis for the origin of BDs) rather difficult if the
evidence is based only on a distributed, binary-deficient BD
population which also has truncated disks when young. It is thus
essential to use the entire information, namely the binary properties
in dependence of the locations and velocities of the BD population.
Thus, the embryo-ejection hypothesis implies a shallow density
distribution of mostly single BDs extending to large distances, while
the standard model implies a density distribution which is similar to
that of the stars with an extended low-density tail of mostly single
BDs.  The fraction of BDs per star as a function of distance from the
aggregate centre (Figs~\ref{fig:rprT0}, \ref{fig:rprT5}) is thus an
observational diagnostic that should help distinguishing the two
models.

\begin{figure}
\begin{center}
\rotatebox{0}{\resizebox{0.5 \textwidth}{!}
{\includegraphics{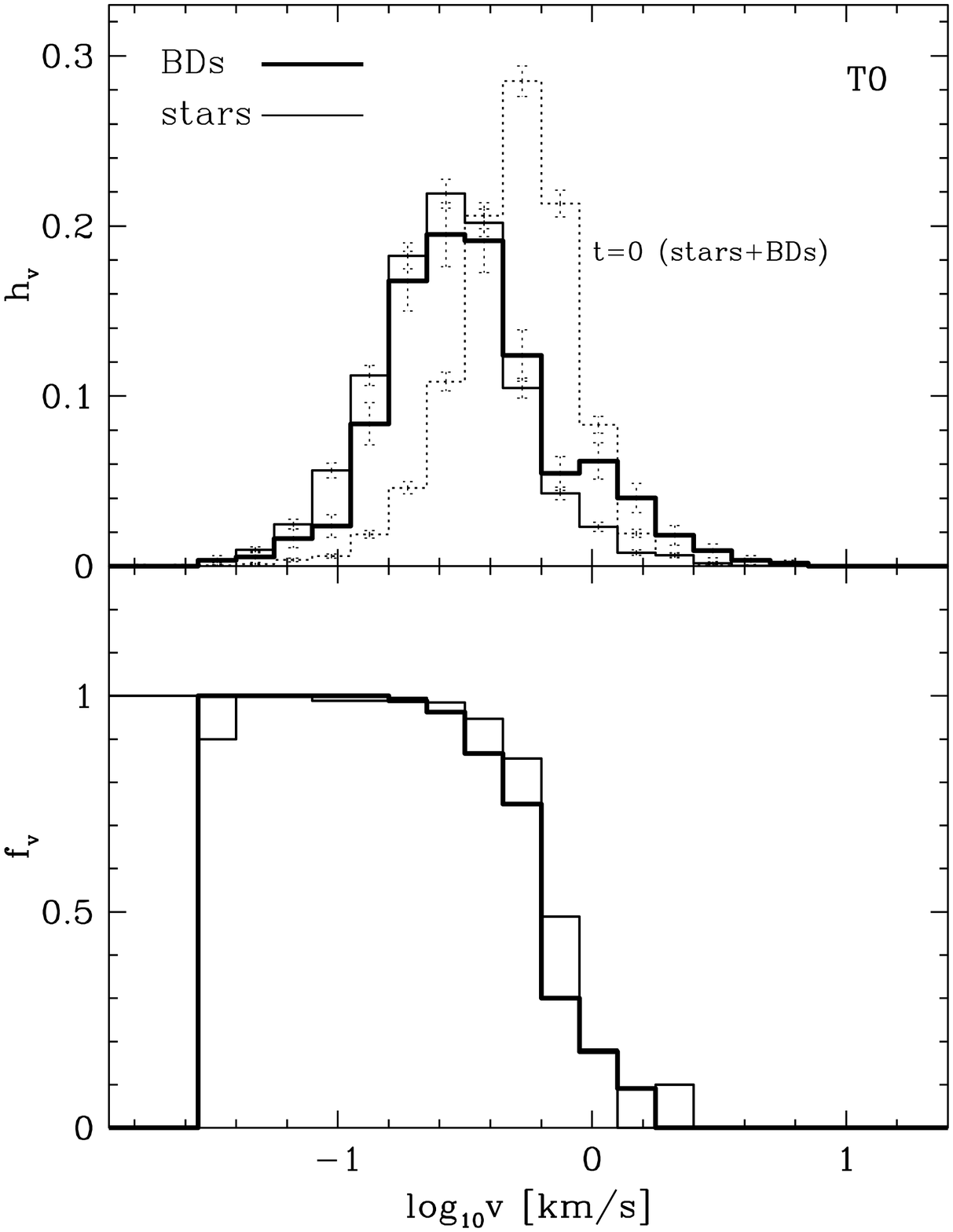}}}
\vskip 0mm
\caption
{{\bf Upper panel:} The average initial velocity distribution of all
stellar and BD systems in model~T0 is shown by the thin dotted
histogram. At late times ($t>10$~Myr) the BD and stellar distributions
have separated. The systems with stellar primaries (thin solid
histogram) still approximately follow a log-normal distribution but
with a reduced mean log$_{10}v$ due to gas removal and expansion of
the aggregate. The BDs (thick solid histogram) have an appreciable
high-velocity (log$_{10}v>1$~km/s) tail which results from the
ejections due to binary--binary encounters.  The velocity distribution
functions are normalised to unit area.  {\bf Lower panel:} The thin
solid histogram is the velocity-dependent binary proportion, $f_v$, of
stellar primaries.  The thick solid histogram is $f_v$ of BD
primaries. In both panels averages of 140~model renditions are
plotted, and the errorbars are standard deviation of the mean values.
}
\label{fig:fv1}
\end{center}
\end{figure}

\begin{figure}
\begin{center}
\rotatebox{0}{\resizebox{0.5 \textwidth}{!}
{\includegraphics{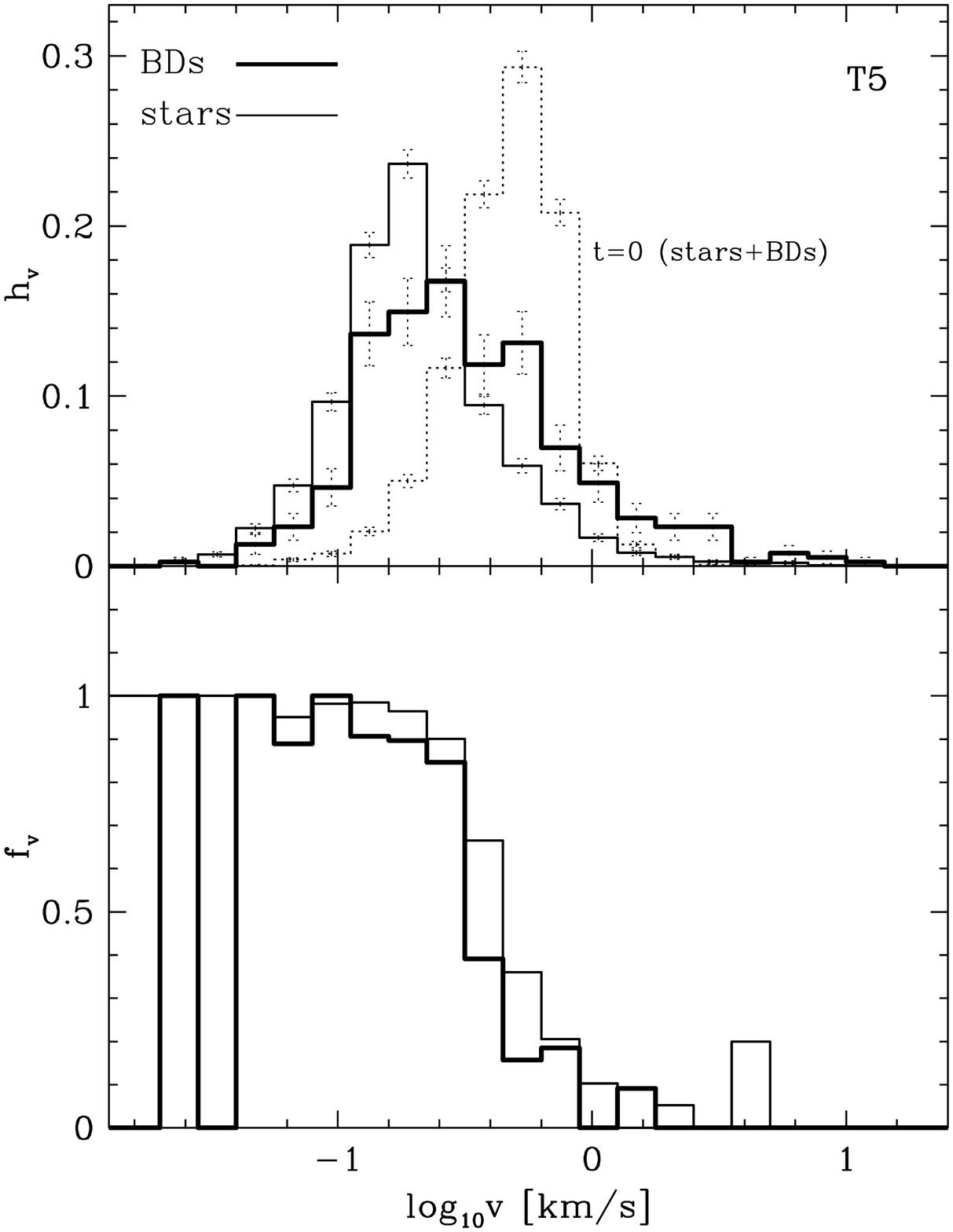}}}
\vskip 0mm
\caption
{As Fig.~\ref{fig:fv1} but showing model~T5.}
\label{fig:fv2}
\end{center}
\end{figure}

The velocity distribution functions are quantified in
Figs.~\ref{fig:fv1} and~\ref{fig:fv2} for models~T0 and~T5,
respectively (the other models T2-T4 yield very similar results, while
model~T1 has a narrower velocity distribution and a much smaller
ejection tail).  Model~T5 enjoys a longer embedded phase which leads
to a more pronounced high-velocity tail for BDs, since the system
remains in a more compact configuration for a longer time allowing
more encounters than in model~T0.

By about 15~Myr about 15~per cent of all BD systems have a velocity
larger than~1~km/s in both models.  The final stellar velocity
distribution is centred on smaller velocities in~T5 than in
model~T0. This results from the slower gas-removal allowing the
stellar orbits to adjust more to the changing conditions than in
model~T0 (T5 is more adiabatic than T0), while the BD population has
more time to be accelerated to higher velocities on average by the
encounters with the stellar members.  Effectively, models~T5 (and
T2--T4) thus lead to a more pronounced kinematical separation between
stars and BDs. This effect will, however, be difficult to observe for
any individual aggregate because of the small number of systems per
aggregate (see Figs.~\ref{fig:app1} and~\ref{fig:app2}).

The figures also show the binary proportion as a function of
velocity. The ejected BDs have a low if not negligible binary
proportion (as for the embryo-ejection hypothesis), and their binary
proportion is systematically smaller in the velocity interval
0.4--1~km/s than that of stellar primaries. The standard model with
BDs implies that the BDs retain a high binary proportion for
$v\simless 0.6$~km/s. Such BDs are spatially distributed similarly to
the stars.  This last point is an important observable diagnostic
distinguishing the embryo-ejection hypothesis from the standard model
with BDs, given that the likelihood for ejection of two embryos that
become a binary BD is exceedingly small.
%=================================== FIGURE ==============================
% /NB_EVAL/Bouv/Taurus/vel_T0.sm
%=========================================================================
%=================================== FIGURE ==============================
% /NB_EVAL/Bouv/Taurus/vel_T5.sm
%=========================================================================

%%%%%%%%%%%%%%%%%%%%%%%%%%%%%%%%%%%%%%%%%%%%%%%%%%%%%%%%%%%%%%%%%%%%%%%%%
\section{Concluding remarks}
\label{sec:concs}

The dynamical evolution of Taurus--Auriga-like populations is studied
using high-precision $N-$body computations of embedded aggregates of
25 binaries with different initial concentrations. The aggregates
remove their gas on time-scales of a fraction to a few crossing
times. The standard model (random pairing from the IMF, no correlation
between binary-star orbital parameters and primary mass, and no
initial mass segregation) is used to define the input population.

The aggregates disperse within about 10~Myr leaving a population that
is dynamically largely unevolved. In particular, the binary population
remains high with $f>0.8$. This shows that for the initial conditions
chosen for the aggregates, the binary destruction by $N-$body
processes is minor and leaves a population that is rich in binaries as
that observed. Taurus--Auriga-type star formation can therefore be
only a minor contributor to the Galactic field population which has
$f=0.55-0.6$.

The distributed BD population has a negligible binary proportion and
truncated disks, as a result of this population component being
preferably ejected during binary--binary encounters from the
aggregates.  This makes it difficult to empirically verify the
hypothesis that BDs may originate as ejected unfinished stellar
embryos from few-body systems using the binary proportion, disk
truncation and wide spatial distribution of BDs as sole
evidence. However, according to the standard model used here (i.e. BDs
form with the same kinematical, spatial and binary properties as
stars), slow BDs which also share a similar spatial distribution as
the stars retain a high binary proportion. This is a powerful
distinguishing diagnostic from the embryo-ejection hypothesis, because
the slow-moving population is much easier to find given that it
overlaps the stellar distribution.

The computations presented here focus on T-A aggregates, but are
equally applicable to other young low-mass populations. For example,
Chauvin et al. (2002) find the $\simless 5$~Myr old MBM12 association
to have an excess binary fraction over Galactic field G~ and
M~dwarfs. This may be explainable through an origin of the MBM12 stars
in low-mass aggregates similar to T-A.

%%%%%%%%%%%%%%%%%%%%%%%%%%%%%%%%%%%%%%%%%%%%%%%%%%%%%%%%%%%%%%%%%%%%%%%%%
\section*{acknowledgements} 
We thank Gaspard Duch\^ene and Estelle Moraux for useful discussions.
PK thanks the staff of the Observatoire de Grenoble for their very
kind hospitality and the Universit\'e Joseph Fourier for supporting a
very enjoyable and productive stay during the summer of 2002.

%%%%%%%%%%%%%%%%%%%%%%%%%%%%%%%%%%%%%%%%%%%%%%%%%%%%%%%%%%%%%%%%%%%%%%%%%

\vfill 


\begin{thebibliography}{}

\bibitem{} Aarseth S.J., H\'enon M., Wielen R., 1974, A\&A, 37, 183

\bibitem{} Bate M.R., 2003, in IAU Symp. 211, Martin E.L. et
	al. (eds.), in press

\bibitem{} Bate M.R., Bonnell I.A., Bromm V., 2002, MNRAS, 332, L65 

\bibitem{} Binney J., Tremaine S., 1987, Galactic Dynamics,
        Princeton University Press, Princeton 

\bibitem{} Bonnell I.A., Davies M.B., 1998, MNRAS, 295, 691

\bibitem{} Bonnell I.A., Bate M.R., Zinnecker H., 1998, MNRAS, 298, 93

\bibitem{} Bontemps S., André P.; Kaas A.A., et al., 2001, A\&A, 372, 173 

\bibitem{} Boss A.P., 2001, ApJ, 551, L167 

\bibitem{} Briceno C., Luhman K.L., Hartmann L., Stauffer J.R.,
	Kirkpatrick J.D., 2002, ApJ, 580, 317

\bibitem{} Cesaroni R., Codella C., Furuya R.S., Testi L., 2003, A\&A,
	in press

\bibitem{} Chabrier G., 2001, ApJ, 554, 1274

\bibitem{} Chauvin G.; M\'enard F.; Fusco T., Lagrange A.-M., Beuzit
	J.-L., Mouillet D., Augereau J.-C., 2002, A\&A, 394, 949

\bibitem{} Clarke C.J., Bouvier J., 2000, MNRAS, 319, 457

\bibitem{} Duch\^ene G., 1999, A\&A, 341, 547 

\bibitem{} Duquennoy A., Mayor M., 1991, A\&A, 248, 485 

\bibitem{} Figueredo E., Blum R.D., Damineli A., Conti P.S., 2002, AJ,
	in press (astro-ph/0204348)

\bibitem{} Geyer M.P., Burkert A., 2001, MNRAS 323, 988

\bibitem{} Gomez M., Hartmann L., Kenyon S.J., Hewett R., 1993, AJ, 105, 1927

\bibitem{} Hartmann L., 2002, ApJ, 578, 914

\bibitem{} Hartmann L., Ballesteros-Paredes J., Bergin E.A., 2001,
	ApJ, 562, 852 

\bibitem{} Hillenbrand L.A., 1997, AJ, 113, 1733

\bibitem{} Hillenbrand L.A., Hartmann L.W., 1998, ApJ, 321, 540 
		
\bibitem{} Hillenbrand L.A., Carpenter J.M., 2000, ApJ, 540, 236

\bibitem{} Klessen R.S., 2001, ApJ., 550, L77

%\bibitem{} Kroupa P., 1995a, MNRAS, 277, 1491 (K1)

\bibitem{} Kroupa P., 1995, MNRAS, 277, 1507 (K2)

%\bibitem{} Kroupa P., 1998, MNRAS, 298, 231

%\bibitem{} Kroupa P., 2000, NewA, 4, 615

\bibitem{} Kroupa P., 2000, in Massive Stellar Clusters, C.M.~Boily,
	A.~Lancon (eds), ASP Conf.~Ser.~211, 233

%\bibitem{} Kroupa P., 2001, IAU Symp., 200, 199 

\bibitem{} Kroupa P., 2001, MNRAS, 322, 231

\bibitem{} Kroupa P., 2002, Science, 295, 82 (astro-ph/0201098)

%\bibitem{} Kroupa P., 2002b, in Modes of Star Formation, E.Grebel,
%W.Brandner (eds), ASP Conf. Ser., in press (astro-ph/0102155)

%\bibitem{} Kroupa P., 2002b, MNRAS, 330, 707

\bibitem{} Kroupa P., Bouvier J., 2003, MNRAS, in prep.

\bibitem{} Kroupa P., Burkert A., 2001, ApJ, 555, 945

\bibitem{} Kroupa P., Aarseth S.J., Hurley J., 2001, MNRAS, 321, 699
	(KAH)

\bibitem{} Kroupa P., Bouvier J., Duch\^ene G., Moraux E., 2003,
	MNRAS, submitted

\bibitem{} Kroupa P., Petr M.G., McCaughrean M.J., 1999, NewA, 4, 495

%\bibitem{} Lada C.J., Margulis M., Dearborn D., 1984, ApJ, 285, 141 (LMD)

\bibitem{} Lada E.A., 1999, in NATO Science Series C Vol. 450, The
	Origins of Stars and Planetary Systems, ed. C.J. Lada \& N.D. Kylafis
	(Dordrecht: Kluwer), 441

%\bibitem{} Lada C.J., Lada E.A., 1991, The Nature, Origin
%        and Evolution of Embedded Star Clusters. In: James K. (ed.), The
%        Formation and Evolution of Star Clusters, ASP Conf. Series 13, p.3

\bibitem{} Malkov O., Zinnecker H., 2001, MNRAS, 321, 149

\bibitem{} Matzner C.D., McKee C.F., 2000, ApJ, 545, 364

%\bibitem{} Moraux E., Kroupa P., Bouvier J., 2002, A\&A, in prep.

\bibitem{} Motte F., Andr\'e P., N\'eri R., 1998, A\&A, 336, 150

\bibitem{} Motte F., Schilke P., Lis D.C., 2002, ApJ, in press 
	(astro-ph/0208519)

\bibitem{} Muench A.A., Lada E.A., Lada C.J., Alves J., 2002, ApJ,
	573, 366

\bibitem{} Muench A.A., Lada E.A., Lada C.J., Elston R.J., Alves J.F.,
	2003, AJ, in press (astro-ph/0301276)

\bibitem{} Palla F., Stahler S.W., 2002, ApJ, submitted (astro-ph/0208554)

\bibitem{} Raboud D., Mermilliod J.-C., 1998, A\&A, 333, 897

\bibitem{} Reipurth B., 2000, AJ, 120, 3177

\bibitem{} Reipurth B., Clarke C., 2001, AJ, 122, 432

\bibitem{} Scally A., Clarke C.J., 2002, MNRAS, 334, 156

\bibitem{} Sterzik M.F., Durisen R.H., 1998, A\&A, 339, 95

\bibitem{} Woitas J., Leinert C., K{\" o}hler R., 2001, A\&A, 376, 982 


\end{thebibliography}
\end{document}